\documentclass[aps,prb,preprint,groupedaddress,showpacs,showkeys,amsmath]{revtex4-1}
\usepackage[dvips]{graphicx}
\usepackage{dcolumn}
\usepackage{bm}
\usepackage{color}
\usepackage{longtable}
\begin{document}

\title{Low-pressure phase diagram of crystalline benzene from quantum Monte Carlo}

\author{Sam Azadi}
\email{s.azadi@ucl.ac.uk}
\affiliation{Department of Physics and Astronomy, University College London,
Thomas Young Center, London Centre for Nanotechnology, 
London  WC1E 6BT, United Kingdom}
\author{R. E. Cohen}
\affiliation{Extreme Materials Initiative, Geophysical Laboratory, 
Carnegie Institution for Science, Washington DC, 20015, USA; 
Department of Earth- and Environmental Sciences, 
Ludwig Maximilians Universit\"{a}t, Munich 80333, Germany;
and Department of Physics and Astronomy, University College London, London  WC1E 6BT, UK }
\date{\today}

\begin{abstract}
We study the low-pressure (0 to 10 GPa) phase diagram  of crystalline benzene 
using quantum Monte Carlo (QMC) and density functional theory (DFT) methods.
We consider the $Pbca$, $P4_32_12$, and $P2_1/c$ structures as the best candidates 
for phase I and phase II. 
We perform diffusion quantum Monte Carlo (DMC)
calculations to obtain accurate static phase diagrams as 
benchmarks for modern van der Waals density functionals.
We use density functional perturbation theory 
to compute phonon contribution in the free-energy calculations.
Our DFT enthalpy-pressure phase diagram indicates that the $Pbca$ and $P2_1/c$ 
structures are the most stable phases within the studied pressure range.  
The DMC Gibbs free-energy calculations predict that the room temperature 
$Pbca$ to $P2_1/c$ phase transition occurs at 2.1(1) GPa.
This prediction is consistent with available experimental results 
at room temperature. Our DMC calculations show an estimate of 
50.6$\pm$0.5 kJ/mol for crystalline benzene lattice energy.  
\end{abstract}
\maketitle

\section {Introduction}

Molecular crystals, including organic and inorganic,
are vital in understanding the
physics and chemistry of the Earth and planets.
They are also of considerable technological interest.
Low-Z molecular systems are among the most abundant
 in the solar system, as represented
by planetary gases and ices. Their behaviour at 
high pressures is crucial in modelling the
structure, dynamic, and evolution of the large planets.
Moreover, compression of molecular systems provides the
opportunities to form new materials, possibly with
novel properties, such as high-temperature superconductivity
and disordered and amorphous materials.
One of the simplest organic molecular solids is crystalline
benzene with aromatic van der Waals (vdW) interactions.
Given its simplicity, high symmetric, and rigid molecular
structure, crystalline benzene has become the model structure
for calculating the lattice model vibrations in molecular crystals.
Benzene has been extensively studied theoretically and experimentally
\cite{Thiery,Wen,Piermarini,Raiteri}.
However, the phase transitions and intermolecular interactions
are still controversial. The main goal of this paper is to present
a comprehensive study of the phase transition of crystalline benzene at
low pressures.

Early experiments by Bridgman\cite{Bridgman}
revealed that liquid benzene crystallises at 68 MPa with $Pbca$
 space group symmetry and closest $C —- C$ intermolecular distance 
of 3.5 $\AA$. This structure, also, was confirmed at 
zero pressure and 270 K\cite{Cox1,Cox2}.
 This phase I is also stable at lower temperatures of 218 and 138 K\cite{Bacon}. 
Since then, two experimental phase diagrams have been proposed for crystalline benzene. 
First, based on the phase diagram suggested by Thi\'{e}ry and L\'{e}ger\cite{Thiery}, 
liquid benzene crystallises at room temperature and pressure 700 bar within an 
orthorhombic structure $Pbca$, which is labeled as phase I. Phase II was suggested 
to exist between 1.4 and 4 GPa. Phases I and II primitive unit cells contain four 
benzene molecules (Z=4). Phase III is stable between 4 and 11 GPa.  
The symmetry of phase III is $P2_1/c$ with two benzene molecules per monoclinic 
primitive unit cell (Z=2). 
Second, the phase diagram developed by Ciabini {\it et al.}\cite{Ciabini1,Ciabini2} which 
based on it phase I is orthorhombic $Pbca$ Z=4 and phase II is monoclinic
 $P2_1/c$ Z=2 \cite{Piermarini}. 
Their results are obtained by means of infrared spectroscopy and X-ray analysis under 
high pressure. The $P2_1/c$ phase is stable up to pressures 20$-$25 GPa. 
This phase diagram only consists of two phases (I and II),
and this same result has been reported by other experiments \cite{Katrusiak2}.
Katrusiak {\it et al}\cite{Katrusiak2},
have determined the crystal structures of phases I and II at 295 K. 
The results of their study confirm Ciabini {\it et al.'s}
phase diagram and show that the structures 
of phases I and II are $Pbca$ Z=4 and $P2_1/c$ Z=2, respectively. The results 
also indicate the absence of other benzene phases in the pressure range up to 5 GPa.

The crystalline benzene phase diagram is a challenge for
first-principles theory because the energy differences are insignificant, 
and they are governed by vdW interactions.
The energy difference between crystalline benzene and its
low-energy polymorphs under pressure is less than few kJ/mol.
Metadynamics calculations predict seven phases\cite{Raiteri} as
phases I ($Pbca$ Z=4), I$^\prime$ ($Cmca$ Z=4), II ($P4_32_12$ Z=4), 
III ($P2_1/c$ Z=2), III$^\prime$ ($C2/c$ Z=4), IV ($Pbam$ Z=4), and 
V ($P2_1$ Z=2). In their calculations, they have used numerous randomly 
generated metastable crystal structures as starting points for the
metadynamics. A few metadynamics steps are often sufficient to obtain
a more stable structure, which most of the time is similar to $Pbca$ Z=4 or 
$P2_1/c$ Z=2.   
Density functional theory (DFT) has also been used to compute the lattice
energy of crystalline benzene\cite{Podeszwa,Grimme}. Wen {\it et al.}
employed DFT formalism and used Perdew-Burke-Ernzerhof (PBE)\cite{PBE}
exchange correlation functionals to study the phase diagram of crystalline 
benzene up to 300 GPa\cite{Wen}. They explained
 the complexities observed in benzene at high pressure. In the 
moderate pressure regime ( P $<$ 20 GPa), they found that the $Pbca$ structure is 
stable up to 4 GPa, the $P4_32_12$ phase is preferred in the pressure 
range of 4$-$7 GPa, and the $P2_1/c$ structure shows the lowest enthalpy at 
higher pressures.  Therefore, they labelled the $Pbca$, $P4_32_12$, and 
$P2_1/c$ structures as phases I, II, and III, respectively. 
The present study shows that the $P4_32_12$ structure is unstable 
in the pressure range of 0$-$10 GPa.
Thus, the $Pbca$ and $P2_1/c$ structures are labelled as
phases I and II, respectively.

Recently, quantum chemistry methods have been 
applied to benzene to obtain sub-kilojoule/mole accuracy in the
lattice energy for crystalline benzene \cite{Yang}. 
Tremendous measures are necessary to obtain such accuracy. 
In this work, we will show that QMC is an alternative efficient
approach to achieve or surpass such accuracy in benzene crystals, 
as we previously demonstrated for the benzene dimer\cite{JCP15}. 

Quantum Monte Carlo (QMC), which approximately 
solves the electronic Schr\"{o}dinger equation
stochastically\cite{Matthew}, can yield highly accurate energies for
atoms\cite{Marchi,Brown}, molecules\cite{Gurtubay,Trail,Azadi2}, 
and crystals\cite{Marchi2,Mostaani,Azadi1}.  
Previous studies have shown that diffusion quantum Monte Carlo (DMC) can
provide accurate energies for vdW systems\cite{Cox,Ma,Yasmine,NABenedek}.
DMC can also produce an accurate description of the phase diagram 
of materials under pressure\cite{Neil,Azadi3,Azadi4}. 
In general, QMC-based methods are faster than the most accurate 
post-Hartree-Fock schemes for large number 
of particles N. The computational cost of QMC methods scales usually
 as $\text N^3$-$\text N^4$ depending on the  method.

We have demonstrated that QMC can provide chemical accuracy for the 
benzene dimer system\cite{JCP15}. We have found optimal 
variational quantum Monte Carlo (VMC) and DMC 
binding energies of $-$2.3(4) and $-$2.7(3) kcal/mol.
The best estimate of the CCSD(T)/CBS limit is $-$2.65(2) kcal/mol\cite{Miliordos}. 
The consistency among our results, experiments, and quantum chemistry
methods, is an important sign of the capability of the QMC-based methods
to provide an accurate description of weak intermolecular interactions
based on vdW dispersive forces. 

In this study, we examine the Z=4 to Z=2 phase transition of crystalline benzene 
at low pressures. We consider the $Pbca$ and $P4_32_12$ structures 
as best candidates for Z=4 and the $P2_1/c$ structure for Z=2. 
We study pressures below 10 GPa. We obtain static and dynamic phase diagrams 
where the phonon contribution to the free energy is included.
We employ different vdW functionals\cite{Klimes2} and 
compare them with conventional DFT functionals. 
We perform QMC calculations to obtain the static enthalpy-pressure phase 
diagram of crystalline benzene. We will show that 
DMC provides accurate results for the phase diagram of crystalline benzene. 

\section {Computational Details}
Given that the energy differences between crystalline benzene 
structures are small, the calculations must be performed with the
highest possible numerical precision.  Our DFT calculations were carried
out within the pseudopotential and plane-wave approach using the Quantum
ESPRESSO suite of programs\cite{QS}. All DFT calculations used
ultrasoft pseudopotentials\cite{ultrasoft}. Pseudopotentials were obtained by
PBE\cite{PBE} exchange correlation functionals.
We used a basis set of plane waves with an energy cutoff 100 Ry. 
Geometry and cell optimisations employed a dense
$12\times12\times12$ ${\bf k}$-point mesh. The quasi-Newton
algorithm was used for cell and geometry optimisation, with convergence
thresholds on the total energy and forces of 0.01 mRy and 0.1 mRy/Bohr,
respectively, to guarantee convergence of the total energy to less
than 1 meV/proton and the pressure to less than 0.1 GPa/proton.

To include the effects of zero point energy (ZPE), 
vibrational frequencies were calculated using density-functional
perturbation theory as implemented in Quantum ESPRESSO\cite{QS}. The ZPE
per proton at a specific cell volume $V$ was estimated within the
quasi-harmonic approximation: $E_{\text{ZPE}}(V) = \hbar
\overline{\omega}/2$, where $\overline{\omega} = \sum_{\bf q}
\sum_{i=1}^{N_{\text{mode}}} \omega_{i}({\bf q})/(N_{\bf q}
N_{\text{mode}})$. $N_{\text{mode}}$ and $N_{\bf q}$ are the
numbers of vibrational modes in the simulation cell and phonon wave vectors
${\bf q}$, respectively, and the summation over ${\bf q}$ includes all
${\bf k}$-points on a $2 \times 2 \times 2$ grid in the Brillouin zone.

The thermodynamic properties are determined by the Helmholtz free energy $F=E-TS$.
The free energy can be written as the sum of an electronic and a vibrational term. 
The electronic entropy is negligible for insulators: $F_{el} \simeq E_{el}$.
In our calculations, the electronic part $E_{el}$ is obtained using the DMC method. 
Thus, the main quantity to calculate for obtaining the thermal properties 
and finite temperature phase diagram is the vibrational free energy $F_{ph}$.
We use quasi-harmonic approximation to calculate the vibrational free energy\cite{Baroni}:

\begin{equation}
  F_{ph} (T,V) = k_{B} T \sum_{i, \bf{q}} ln \{ 2 sinh[\hbar \omega_{i, \bf{q}} (V)/2k_{B}T] \},
\label{eq2}
\end{equation}

where $k_B$, $V$, and $\omega_{i, \bf{q}}$ are Boltzmann constant, unit cell 
volume, and eigenvalue of the phonon Hamiltonian, respectively.
The pressures $P$ are calculated from the Helmholtz free energies by
$P=-(\partial F/ \partial V)_T$

We used the \textsc{casino} code\cite{casino} to perform fixed-node
DMC simulations with a trial wave function of the Slater-Jastrow (SJ)
form:
\begin{equation}
  \Psi_{\rm SJ}({\bf R}) = \exp[J({\bf R})] \det[\psi_{n}({\bf r}_i^{\uparrow})] 
  \det[\psi_{n}({\bf r}_j^{\downarrow})],
\label{eq1}
\end{equation}
where ${\bf R}$ is a $3N$-dimensional vector of the positions of the
$N$ electrons, ${\bf r}_i^{\uparrow}$ is the position of the $i$'th
spin-up electron, ${\bf r}_j^{\downarrow}$ is the position of the
$j$'th spin-down electron, $\exp[J({\bf R})]$ is a Jastrow factor,
and $\det[\psi_{n}({\bf r}_i^{\uparrow})]$ and $\det[\psi_{n}({\bf
  r}_j^{\downarrow})]$ are Slater determinants of spin-up and
spin-down one-electron orbitals.  These orbitals were obtained from
DFT calculations performed with the plane-wave-based Quantum ESPRESSO
code\cite{QS}, employing Trail-Needs\cite{TN1,TN2}
Hartree-Fock pseudopotentials. For the QMC study of C and CH-based systems, 
the Hartree-Fock description of the core is more accurate\cite{Greeff}.
A detailed study of silicon also showed\cite{JMZuo} that Hartree-Fock 
provides the most accurate description of the core density compared with
generalised gradient approximation and local density approximation
(LDA).

We selected a very large basis-set energy cutoff of 200 Ry to approach
the complete basis-set limit\cite{sam}.
The plane-wave orbitals were transformed into a localised ``blip''
polynomial basis\cite{blip}.  Our Jastrow factor
consists of polynomial one-body electron-nucleus, two-body
electron-electron, and three-body  electron-electron-nucleus terms,
the parameters of which were optimised by
minimising the variance of the local energy at the VMC level\cite{varmin1,varmin2}.
Our DMC calculations were performed at two different time steps 0.01 and 0.02 a.u. 
The target population control is two times larger for time step 0.02 a.u. 
We extrapolated our DMC energies to zero time step using a linear fitting. 
The time step error is linear in the time step. 
The population control error also is linear as function of reciprocal of 
the target population. Therefore, it is possible to remove both time 
step and population control errors simultaneously by linearly extrapolation
to zero-time step. 
 
\section {Results and discussion}
\subsection{Geometry Analysis}
In this section we discuss the results of our geometry optimization. 
We study the evolution of benzene molecule distances by increasing the pressure. 
The structure optimization results are compared with experiments. 

\begin{figure}
\begin{tabular}{c c c}
\includegraphics[width=0.25\textwidth]{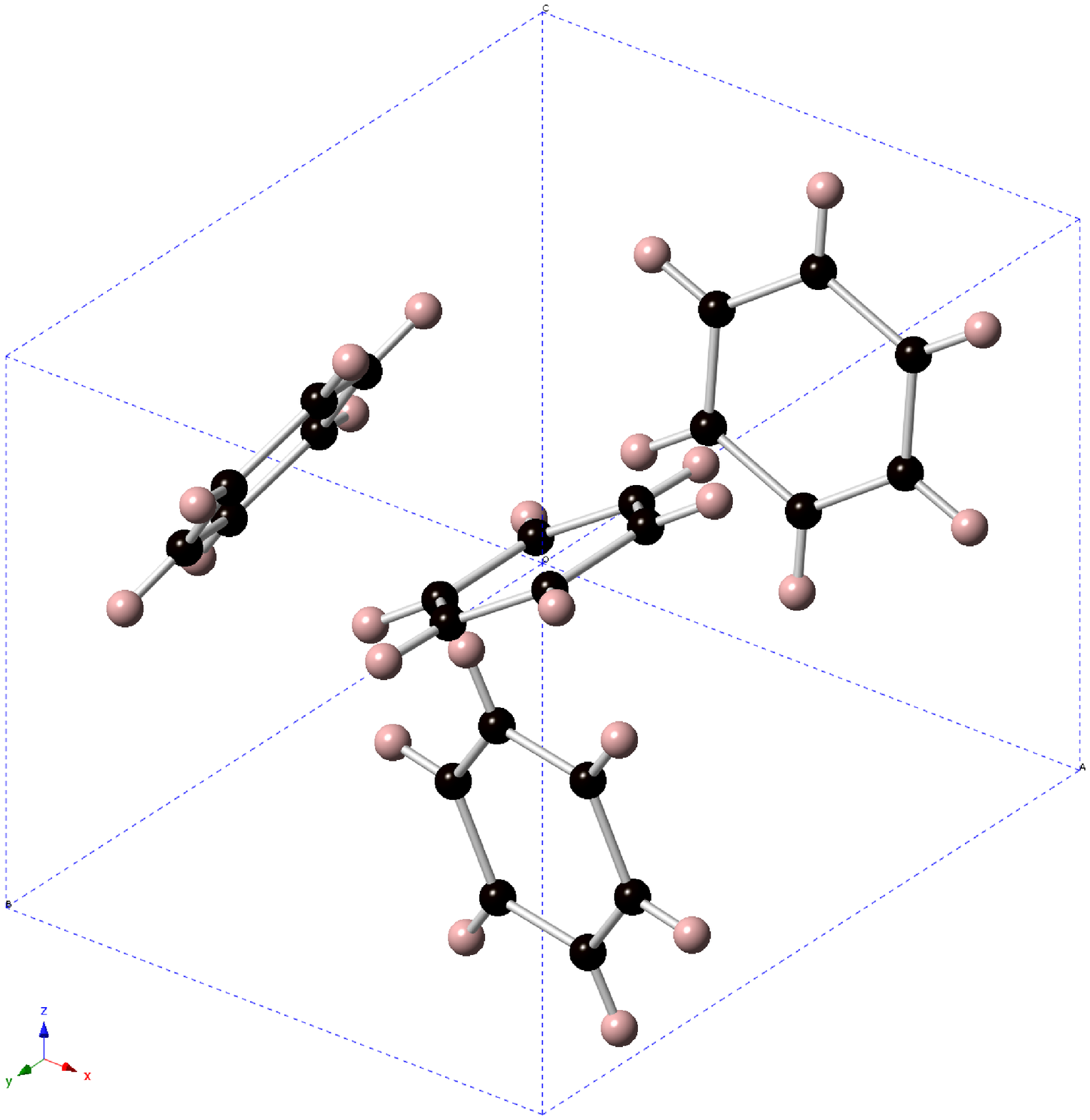} &
\includegraphics[width=0.25\textwidth]{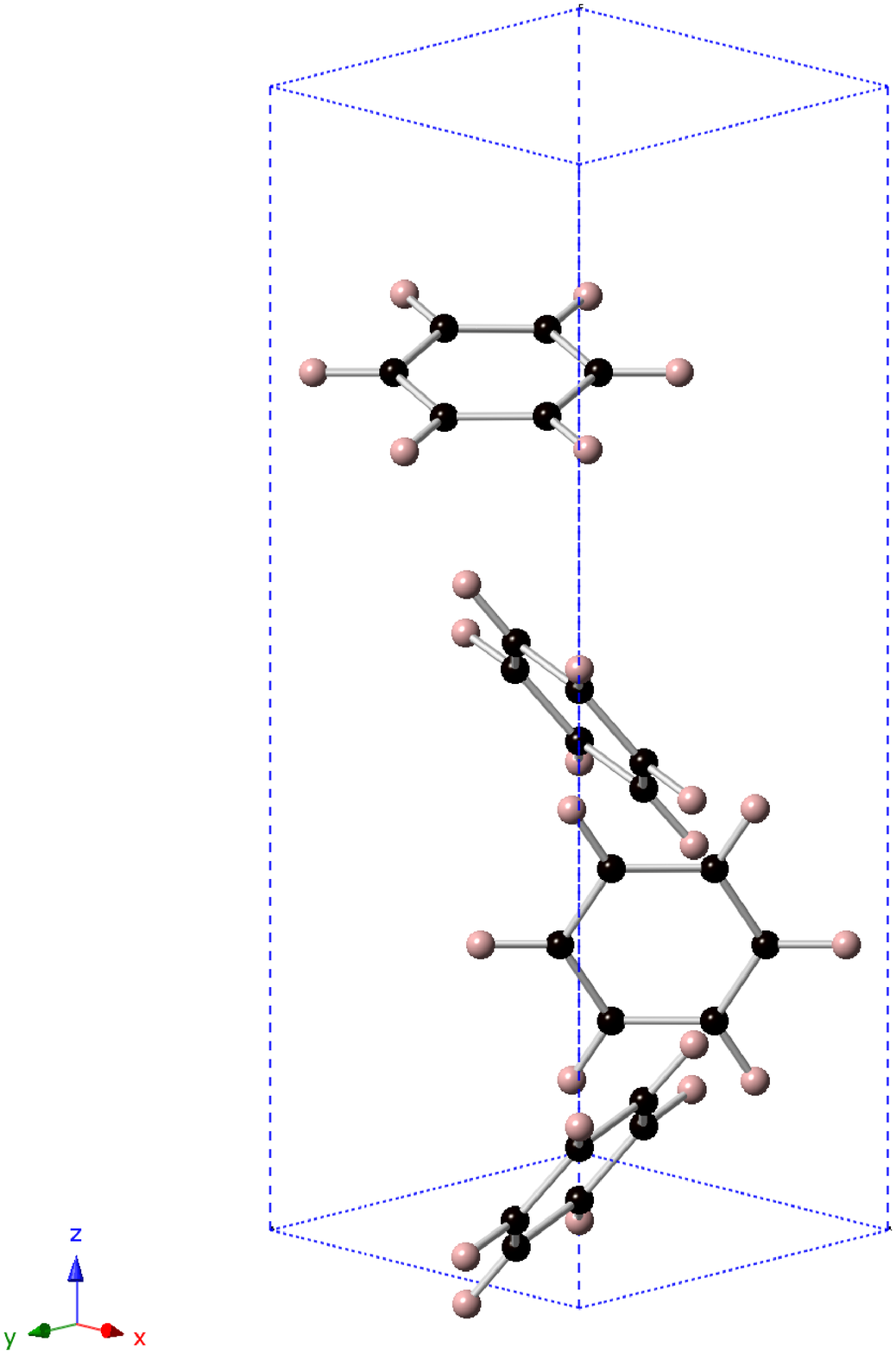} & 
\includegraphics[width=0.25\textwidth]{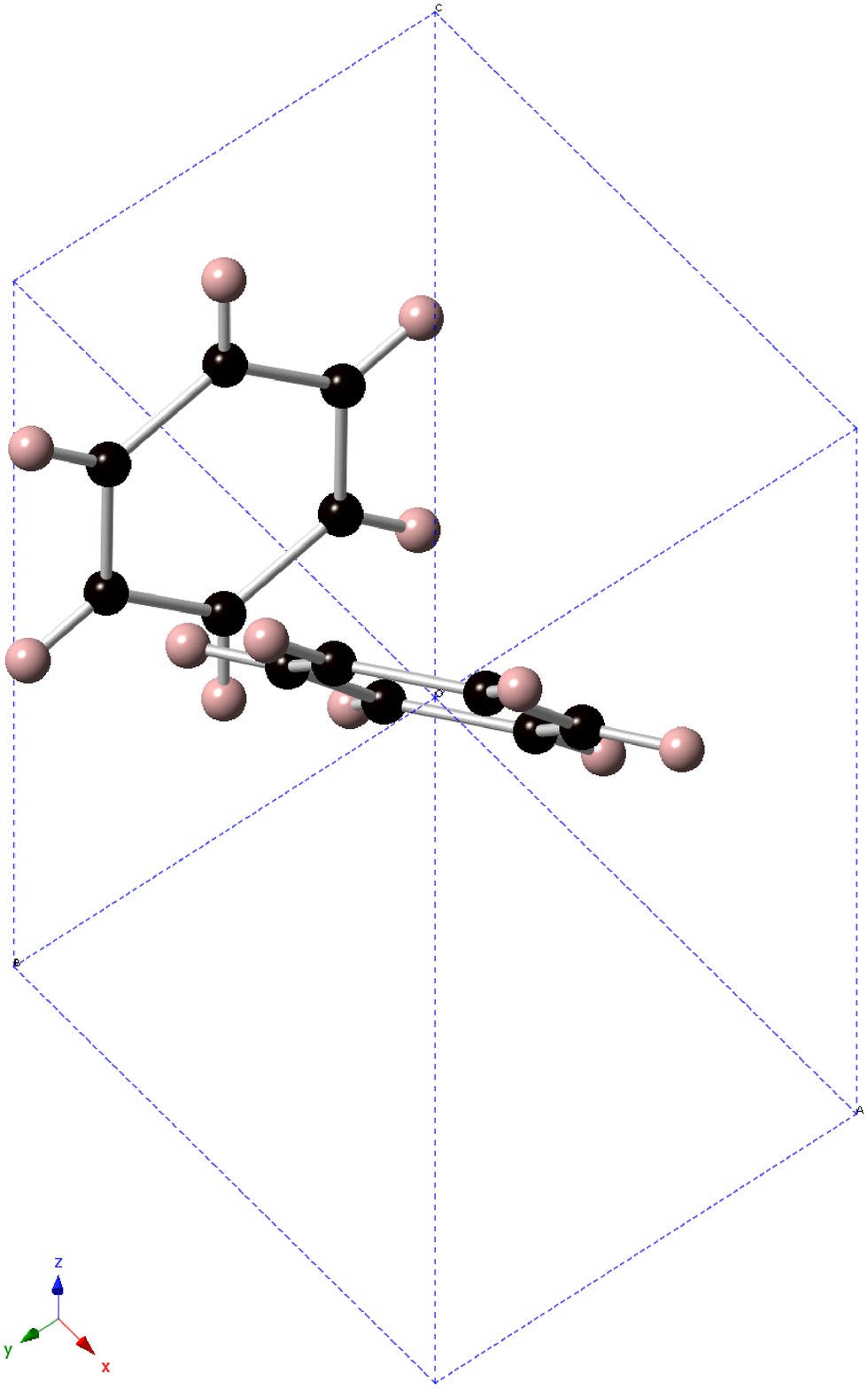}
\end{tabular}
\caption{\label{structures}(colour online) Primitive unit cells 
of the $Pbca$ (left), $P4_32_12$ (middle), and $P2_1/c$ (right) structures of solid benzene
at low-pressure range. The $Pbca$ and $P4_32_12$ primitive unit cells 
contain four  benzene molecules (Z=4), whereas the $P2_1/c$ structure has two benzene 
molecules (Z=2) in a monoclinic primitive unit cell.}
\end{figure}

The primitive unit cells of the $Pbca$, $P4_32_12$, and $P2_1/c$ structures of solid 
benzene contain four, four, and two benzene molecules, respectively, as 
shown in figure~\ref{structures}. The $Pbca$ and $P4_32_12$ structures 
have orthorhombic and tetragonal primitive unit cells, respectively, 
whereas the $P2_1/c$ primitive unit cell is monoclinic. 

\begin{figure}
\begin{tabular}{c c}
\includegraphics[width=0.5\textwidth]{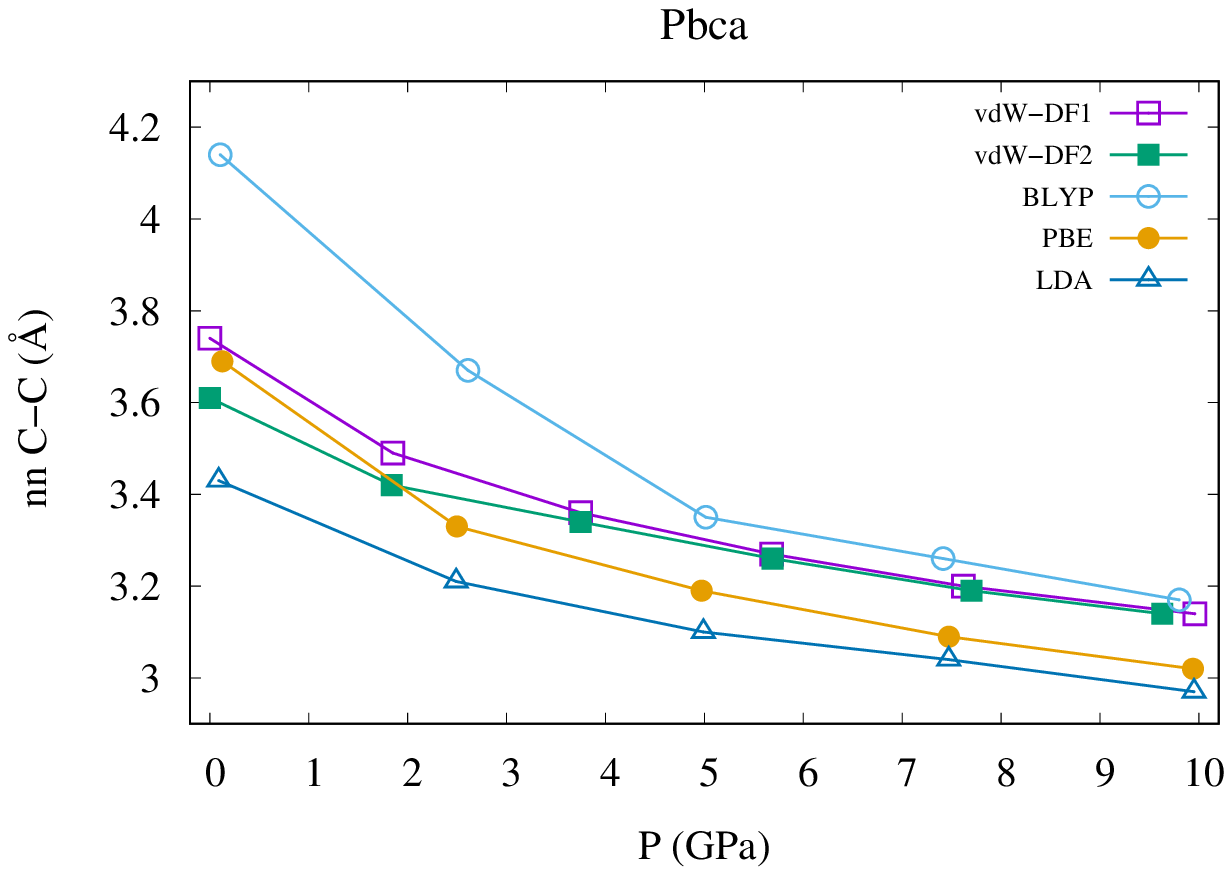}&
\includegraphics[width=0.5\textwidth]{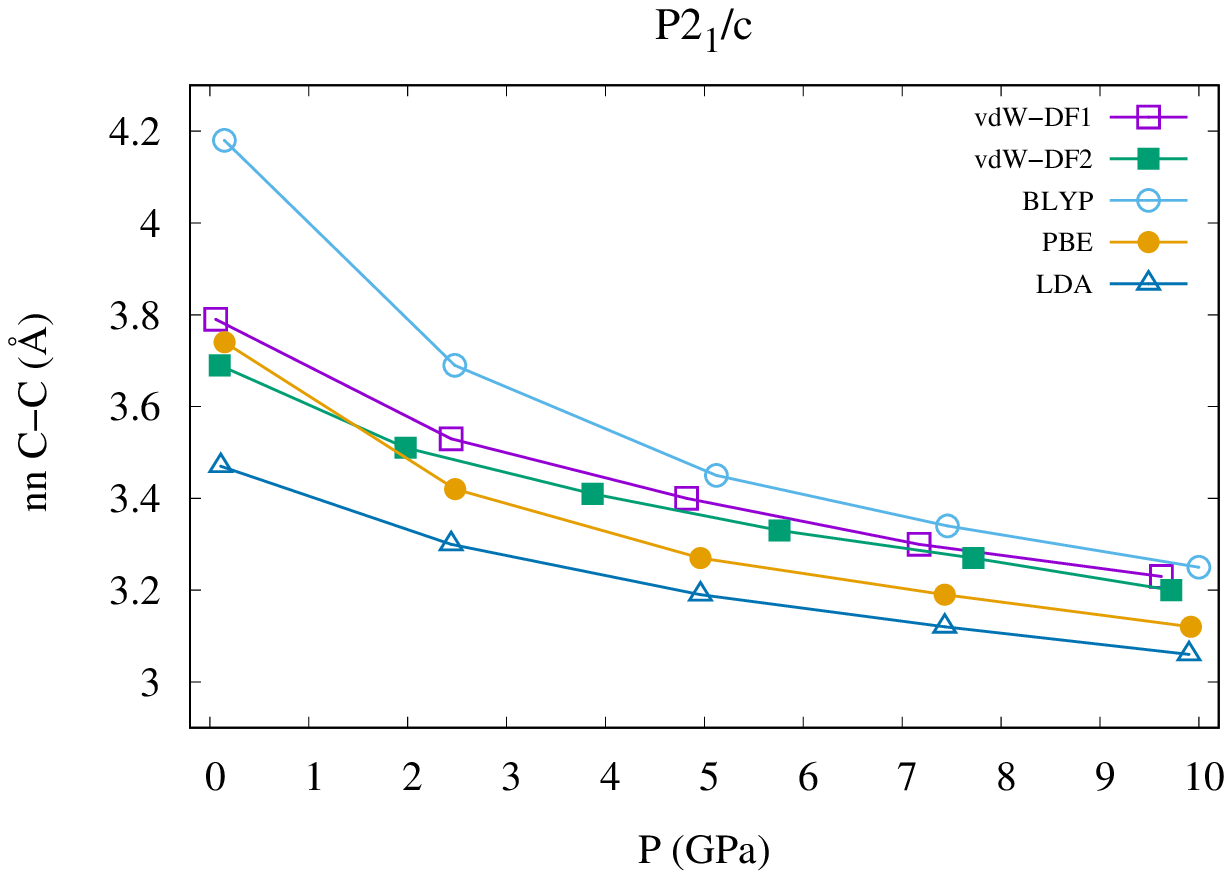} \\
\end{tabular}
\caption{\label{NN}(colour online) Pressure evolution of nearest-neighbour (nn)
C$-$C distances for $Pbca$ and $P2_1/c$. The results are obtained by 
vdW and conventional functionals.} 
\end{figure}

For geometry analysis of Z=4 and Z=2 structures, 
we focus on the $Pbca$ and $P2_1/c$. We will show in the next 
section that these two structures are the best candidates 
for the phases I (Z=4) and II (Z=2).
Our structure optimization indicates that the molecular orientations
do not change significantly within the studied pressure range.
We calculated the distances between C atoms on 
nearest-neighbour (nn) benzene molecules. The nn C$-$C distances between molecules 
as function of pressure are reported in figure~\ref{NN}. The nn C$-$C distances 
for $Pbca$ and $P2_1/c$ structures are calculated using vdW\cite{vdW1,vdW2} and conventional 
DFT functionals.  

The vdW functionals, particularly vdW-DF2\cite{vdW2}, nn C$-$C distances are in good agreement with 
experiment\cite{Ciabini1, Ciabini2}. The differences 
between vdW-DF1\cite{vdW1} and vdW-DF2\cite{vdW2} nn C$-$C distances reduce with increasing pressure. 
The PBE nn C$-$C distances are close to vdW functional results at lower pressures, 
whereas the differences between PBE and vdW results increases with increasing pressure.
The PBE nn C$-$C distances at higher pressures are close to LDA results. 
The BLYP nn C$-$C distances are the largest at low pressures. However, 
BLYP nn C$-$C distances are more similar to vdW results at pressures larger than 5 GPa.  

\begin{figure}
\begin{tabular}{c c}
\includegraphics[width=0.5\textwidth]{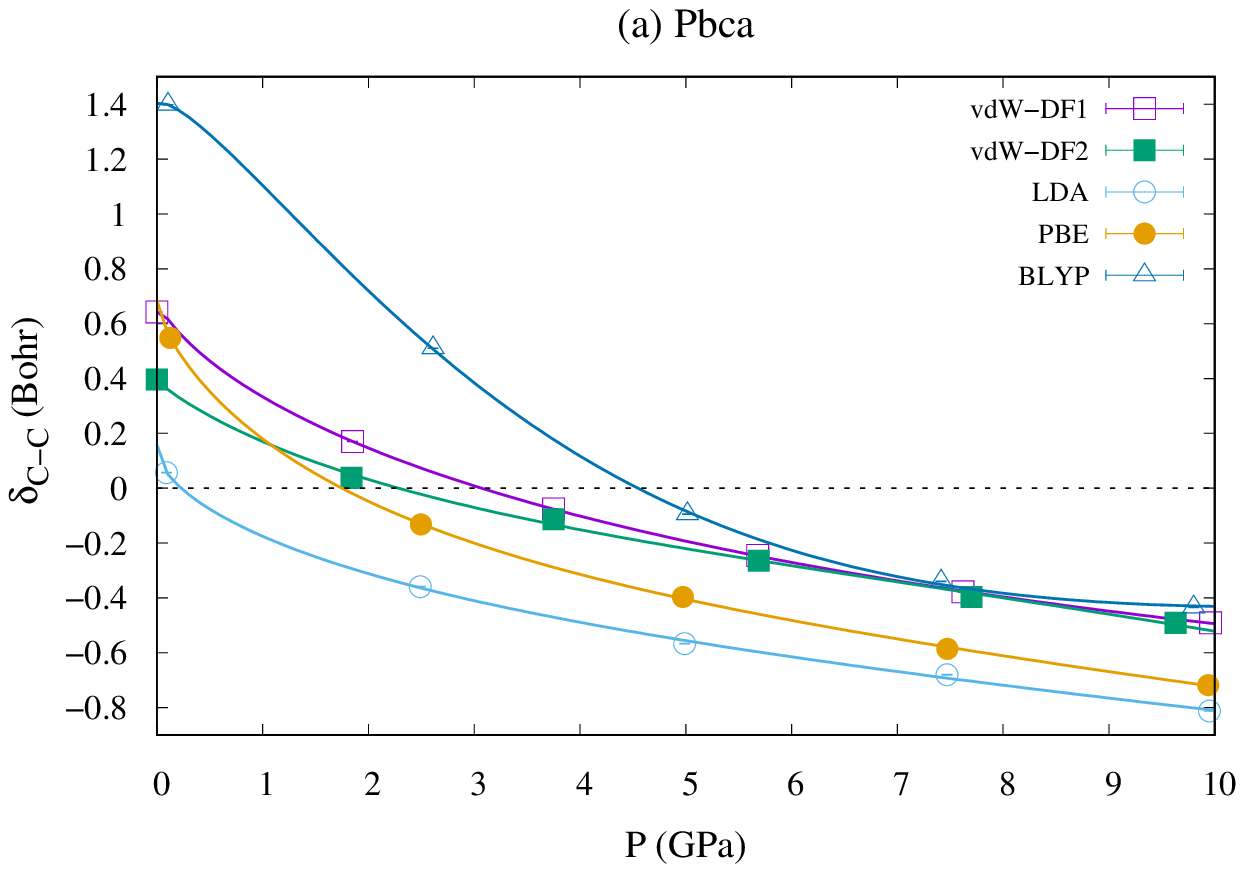}&
\includegraphics[width=0.5\textwidth]{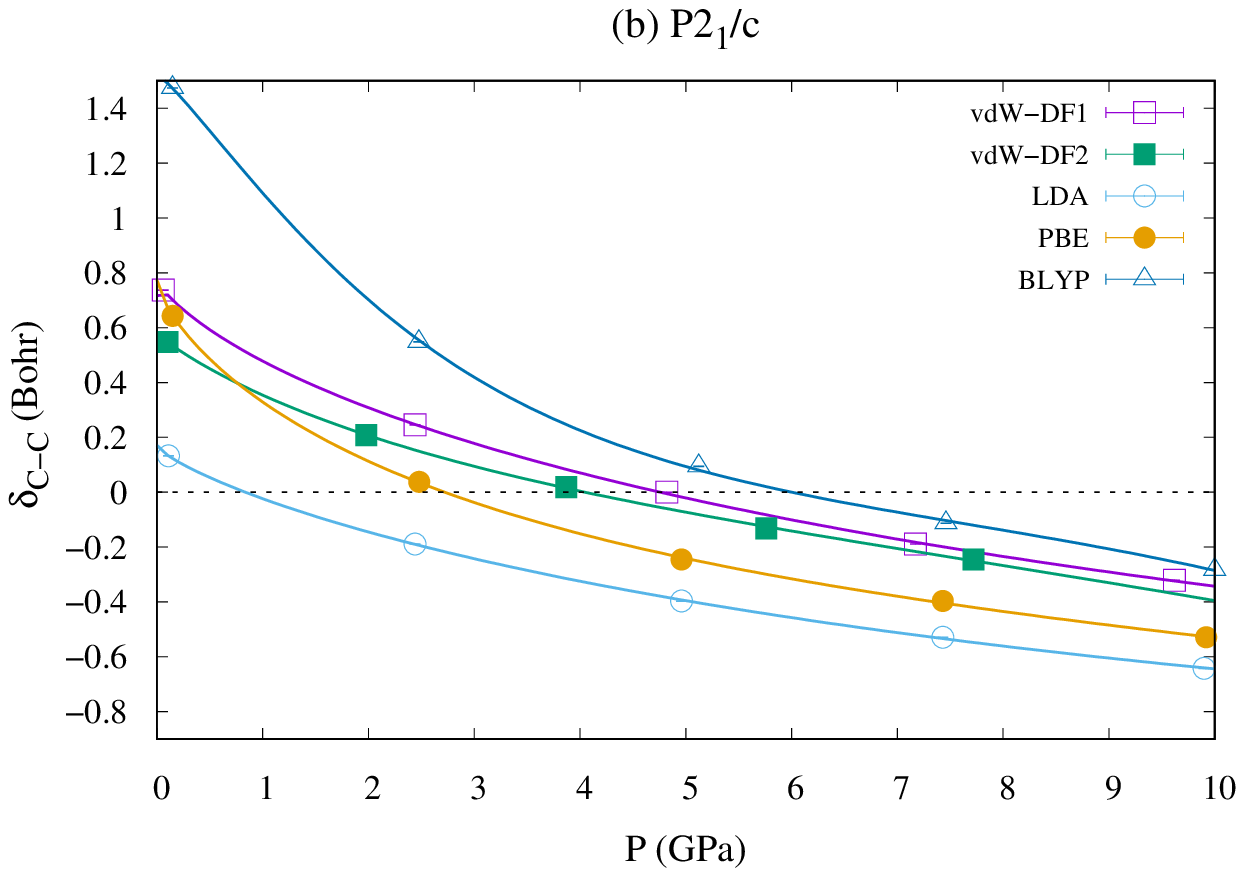} \\
\end{tabular}
\caption{\label{nn_CC}(colour online) Pressure evolution of the difference between nearest-neighbour (nn)
C$-$C distances and the sum of van der Waals radii of C atoms ($\delta_{C-C}$). 
The results are obtained for $Pbca$ and $P2_1/c$ structures using vdW and conventional functionals.} 
\end{figure}

The van der Waals radius of Carbon atom $r_{WC}$ is 1.7 $\AA$.
In crystalline benzene the 
benzene molecules are held together by van der Waals forces.
The nearest that two $C$ atoms belonging to different benzene molecules can approach
each other can be estimated by the sum of $r_{WC}$. 
We calculated the difference between nearest-neighbour (nn)
C$-$C distances and the sum of $r_{WC}$ ($\delta_{C-C}$).  
Figure ~\ref{nn_CC} illustrates $\delta_{C-C}$ for $Pbca$ and $P2_1/c$ structures. 
The results are obtained by vdW and conventional functionals. 
At the same pressure, all the functionals give larger $\delta_{C-C}$ for $P2_1/c$ structure.
Our EOS calculations, which are presented in figure~\ref{PV}, indicate that at the same pressure 
molecular density of $P2_1/c$ is larger than $Pbca$. 
LDA and BLYP provide the smallest and largest $\delta_{C-C}$. Consequently they yield 
the smallest and largest vdW radii for C atom. Unlike the other functionals, 
the BLYP $\delta_{C-C}$ decline rapidly with increasing the pressure.
According to LDA, $Pbca$ results,
benzene molecules are strongly bonded at pressures larger than 0.2 GPa.
In lower pressures PBE $\delta_{C-C}$
is close to $\delta_{C-C}$ obtained by vdW functionals. With increasing the pressure 
PBE results become closer to LDA. According to vdW-DF1 results, the benzene molecules in 
$Pbca$ structure is bonded above 3 GPa. Whereas vdW-DF2 results indicate that bonding between 
benzene molecules in $Pbca$ phase could happen around 2 GPa. Based on 
the experimental phase digram\cite{Ciabini1,Ciabini2}, the $Pbca$ phase 
is stable at pressures below 1.4 GPa. Our vdW $\delta_{C-C}$ results show that there 
are no strong bonds between benzene molecules in $Pbca$ phase. In the $Pbca$ structure 
the benzene molecules only interact through weak dispersive forces.  
  
\subsection{Ground State DFT Phase Diagram}

We begin our phase diagram study by DFT enthalpy-pressure calculations at zero 
temperature. We first present our static phase diagram results where the 
Born–Oppenheimer (BO) approximation is used. According to BO approximation the electronic 
and nuclear wave functions can be separated. It is also assumed that the nuclei are
infinitely massive and the total nuclear momentum contribution in the Hamiltonian is zero. 
To find out the best candidate for Z=4 at the studied pressures, 
we used the PBE\cite{pbesol} and vdW-DF2\cite{vdW2} functionals to calculate 
the enthalpy difference between the $Pbca$, $P4_32_12$, and $P2_1/c$ structures. 
We performed calculations at six different volumes corresponding to 
DFT pressures of 0, 2, 4, 6, 8, and 10 GPa (Figure~\ref{PV}). 
Based on the linear fitting of the PBE results on two enthalpy-pressure points 
at P = 0 and 10 GPa, the $Pbca$ structure is stable up to 3.6 GPa, whereas $P4_32_12$ is 
stable in the pressure range of 3.6$-$6.8 GPa, and finally the $P2_1/c$ structure has 
lowest enthalpy in pressures higher than 6.8 GPa.

A line between these two enthalpy-pressure points gives excellent agreement
with the previous PBE computations by  Wen {\it et al.}\cite{Wen} (Fig. 2(a)).
However we find this result to be inaccurate, and a denser set of points in this
pressure range is needed.

\begin{figure}
\begin{tabular}{c c}
\includegraphics[width=0.5\textwidth]{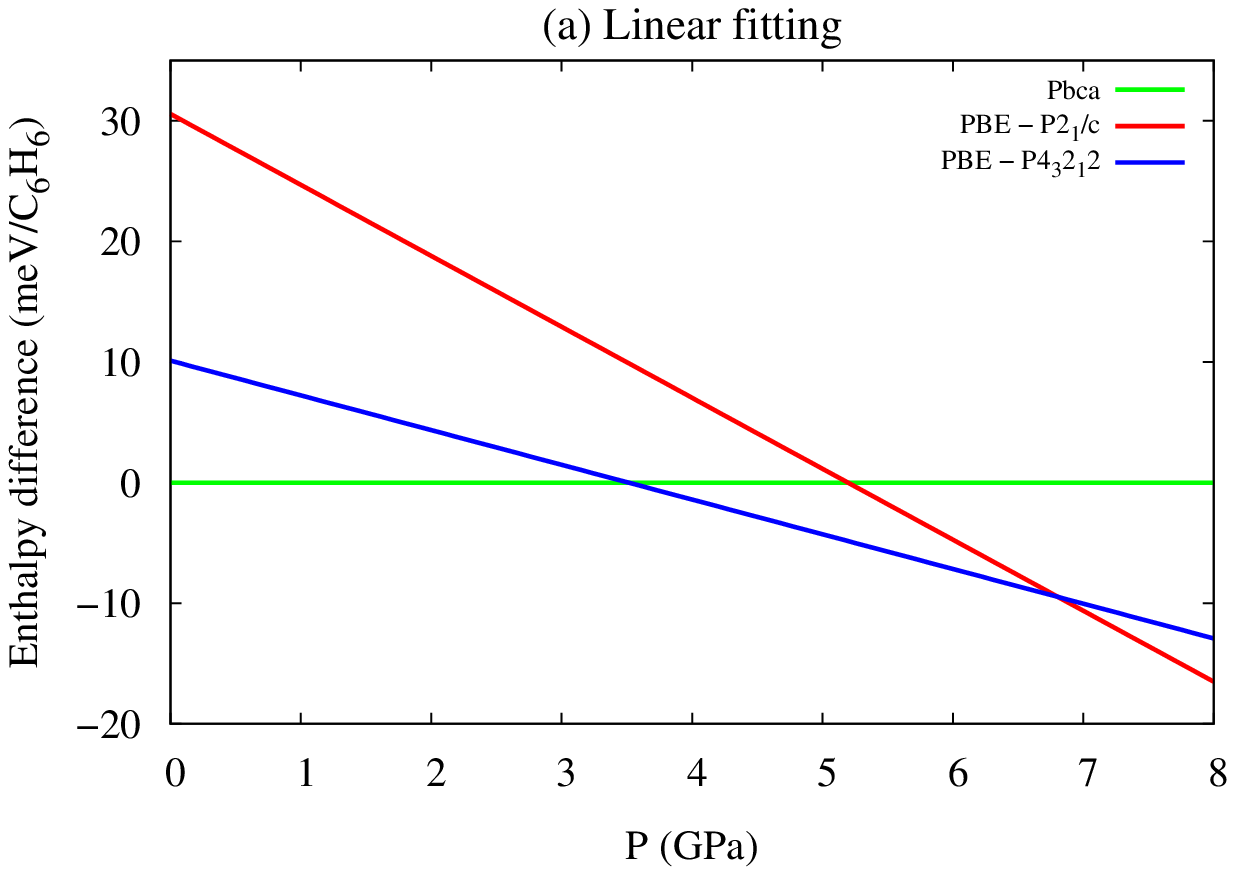} &
\includegraphics[width=0.5\textwidth]{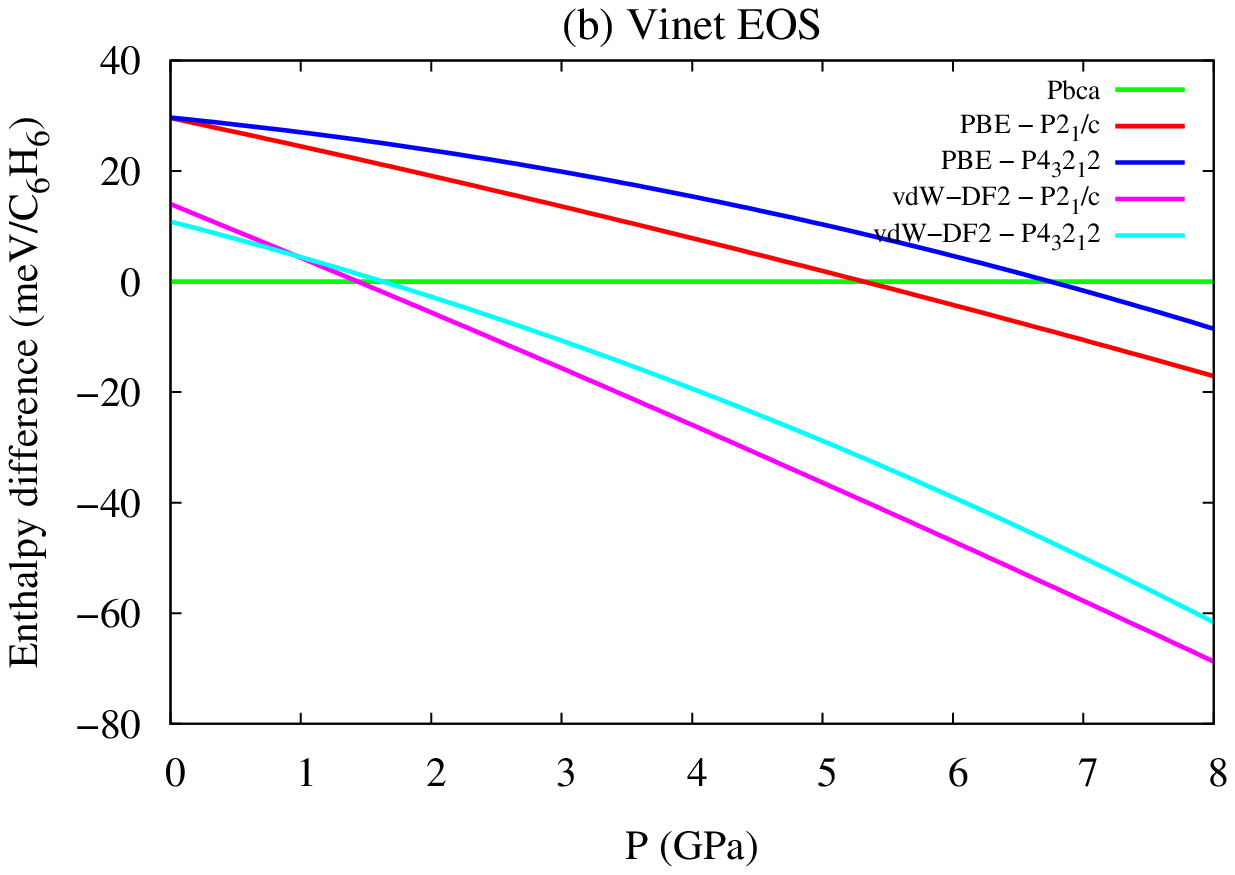}
\end{tabular}
\caption{\label{123} (colour online) Enthalpy difference between the $Pbca$, 
$P4_32_12$, and $P2_1/c$ structures as function of applied pressure.
(a) The results are calculated using PBE and linear fitting on two 
enthalpy-pressure points at P = 0, and 10 GPa. (b) The phase diagram 
is simulated  using DFT-PBE and vdW
density functional of vdW-DF2. We used the Vinet EOS
and six enthalpy-pressure points at P = 0, 2, 4, 6, 8, and 10 GPa.}
\end{figure}

Using the Vinet\cite{Vinet} equation of state (EOS) 
we found that the $P4_32_12$ structure is not stable
in the pressure range of 0$-$10 GPa. The results of our EOS 
calculations are presented in figure~\ref{PV}. 
The enthalpy difference between the $Pbca$, $P4_32_12$, and $P2_1/c$
structures versus pressure is calculated using 
PBE and vdW functionals (Figure~\ref{123}(b)).
We find that instability of $P4_32_12$ is independent of employed functional. 
Our results indicate that $Pbca$ and $P2_1/c$ are the most stable structures in the 
studied pressure ranges. These results are consistent with the experimental phase 
diagram proposed by Ciabini {\it et al.}\cite{Ciabini1,Ciabini2,Piermarini,Katrusiak2}. 
Therefore, in the rest of this paper, we label $Pbca$ and $P2_1/c$ as phases I 
and II, respectively.

\begin{figure}
\begin{tabular}{c c}
\includegraphics[width=0.5\textwidth]{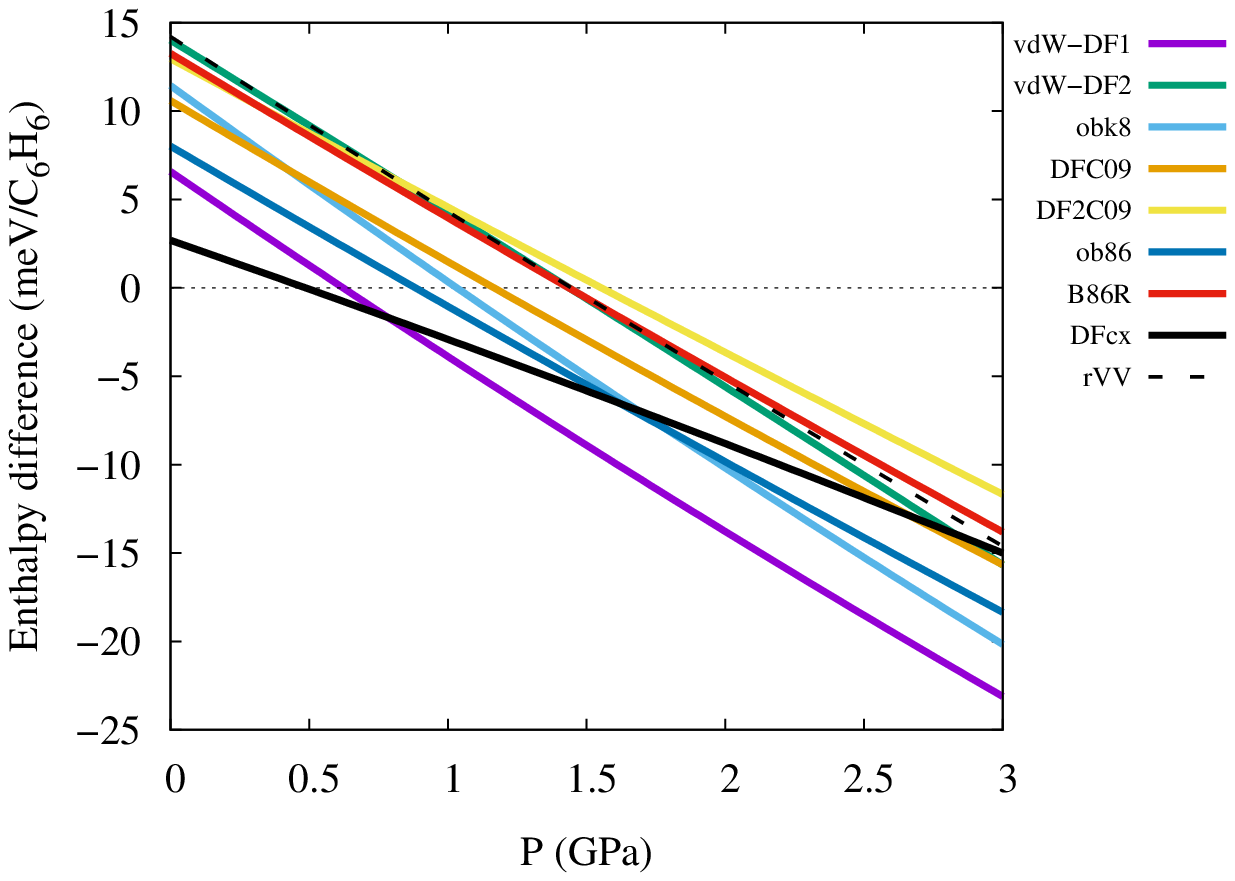}& 
\includegraphics[width=0.5\textwidth]{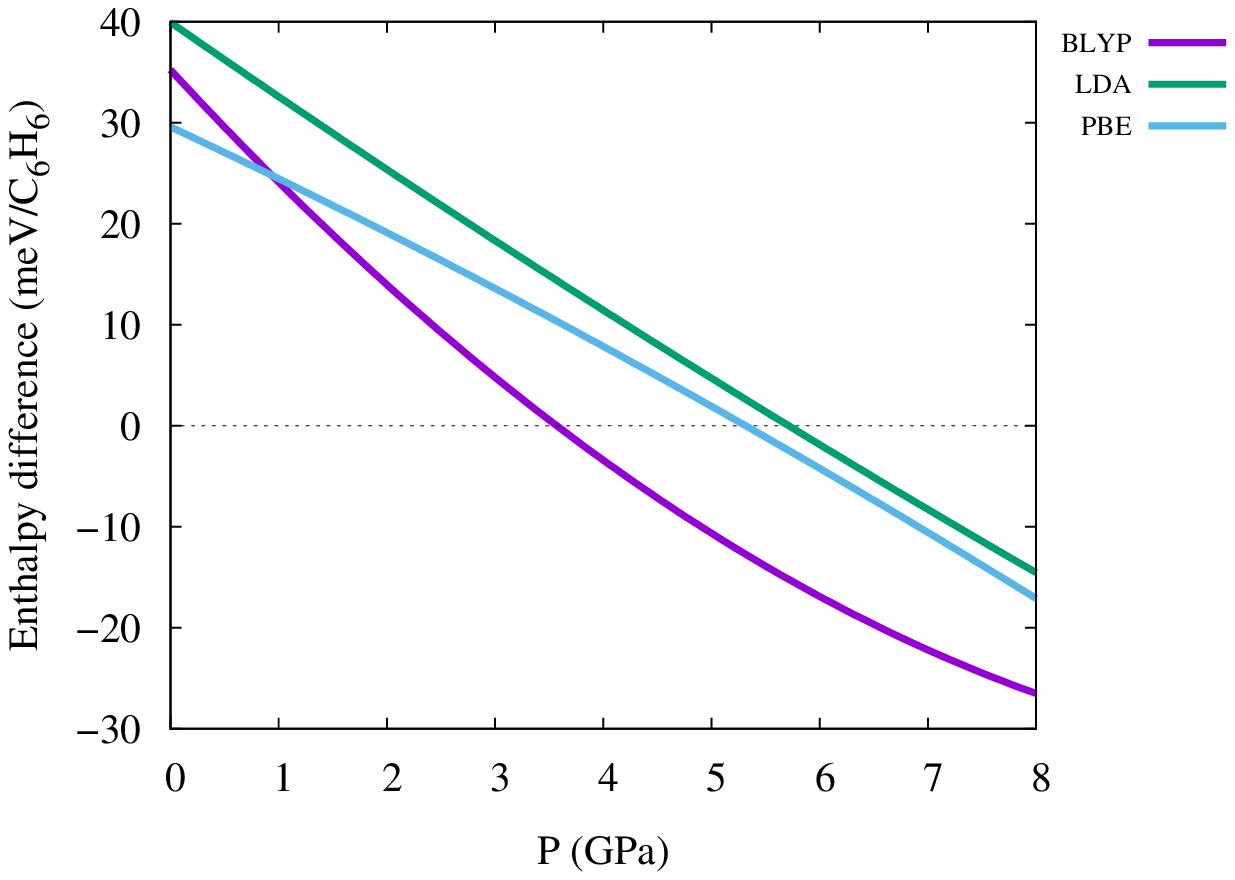}\\ 
\end{tabular}
\caption{\label{DFT_enthalpy}(colour online) Enthalpy difference
between $Pbca$ (phase I) and $P2_1/c$ (phase II) as function of
pressure obtained with vdW and conventional DFT funcationals.
The left panel shows the results of vdW-DF1\cite{vdW1}, vdW-DF2\cite{vdW2},
rVV\cite{rVV1,rVV2}, obk8, B86R, ob86\cite{Klimes1,Klimes2}, 
DFCx\cite{cx}, DFC09, and DF2C09\cite{C09} vdW functionals.
The right panel illustrates the results of conventional DFT 
including PBE\cite{pbesol}, LDA\cite{LDA}, and BLYP\cite{BLYP}.} 
\end{figure}

To study the importance of dispersion effects, we calculated
the phase diagram of crystalline benzene using
different functionals (Figure~\ref{DFT_enthalpy}).
We employed vdW-DF1\cite{vdW1}, vdW-DF2\cite{vdW2},
vdW-DF-obk8, vdW-DF-ob86, vdW-DF2-B86R\cite{Klimes1,Klimes2}, vdW-DF-C09,
vdW-DF2-C09\cite{C09}, vdW-DF-cx\cite{cx}, and 
vdW-rVV\cite{rVV1,rVV2} vdW functionals.
Except rVV functional, the nonlocal term in the other vdW functionals is either 
vdW-DF1\cite{vdW1} or vdW-DF2\cite{vdW2}. 
Employing various gradient corrections
to the exchange energy results in a variety of vdW functionals.
We also determined the phase diagram 
using conventional DFT functionals, including PBE\cite{pbesol},
LDA\cite{LDA}, and BLYP\cite{BLYP}. 

\begin{figure}
\includegraphics[width=0.5\textwidth]{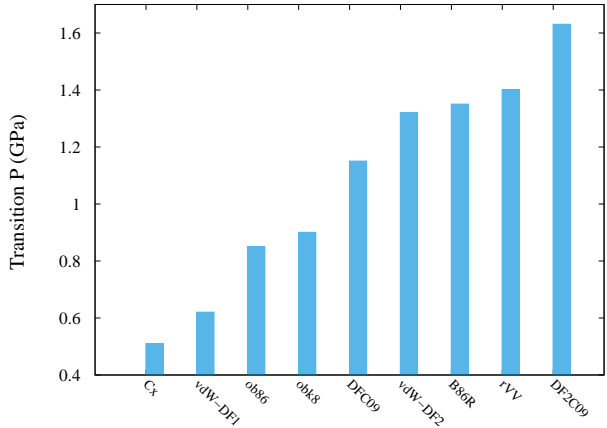}
\caption{\label{Ptran}(colour online) $Pbca$ to $P2_1/c$ phase transition
pressure. The results are calculated using different vdW functionals as 
explained in text. } 
\end{figure}

\begin{figure}
\begin{tabular}{c c}
\includegraphics[width=0.5\textwidth]{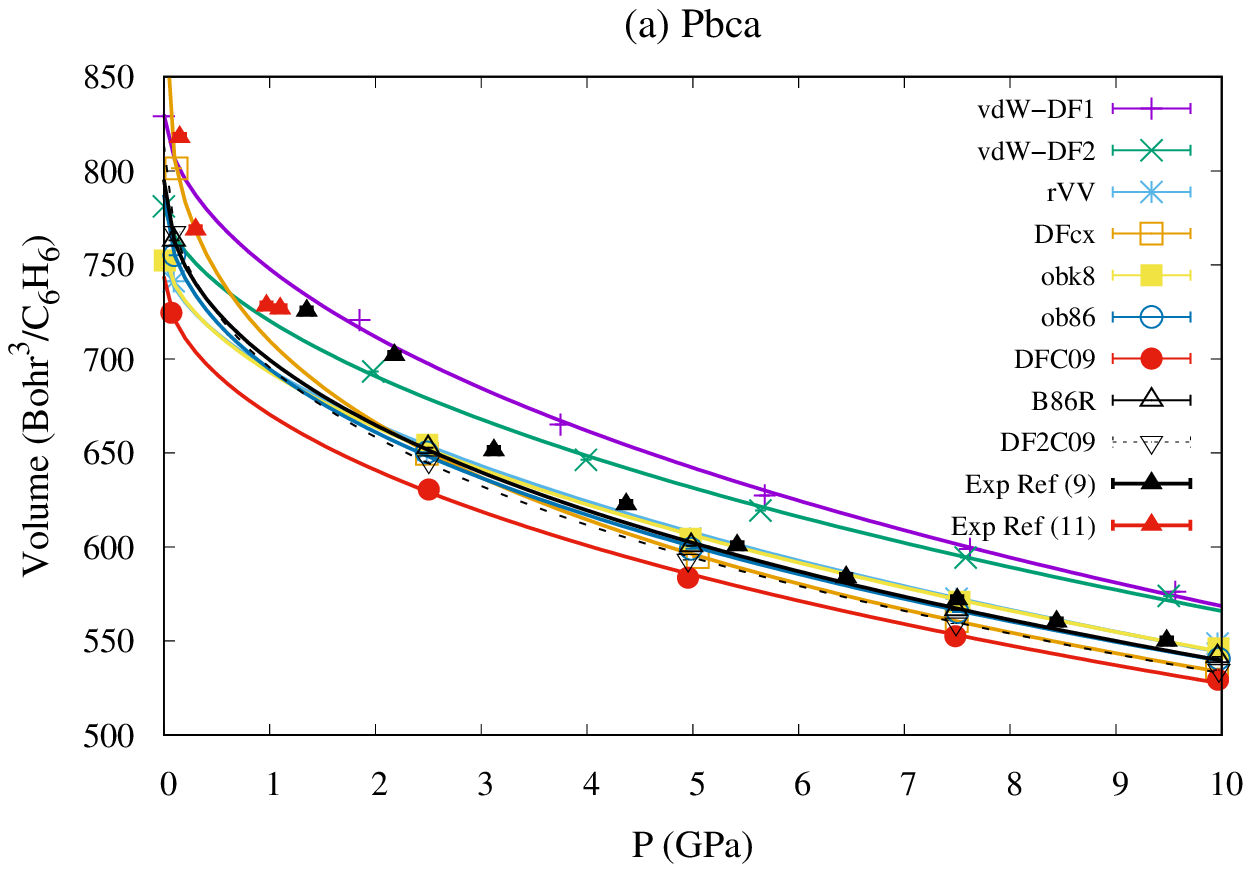}&
\includegraphics[width=0.5\textwidth]{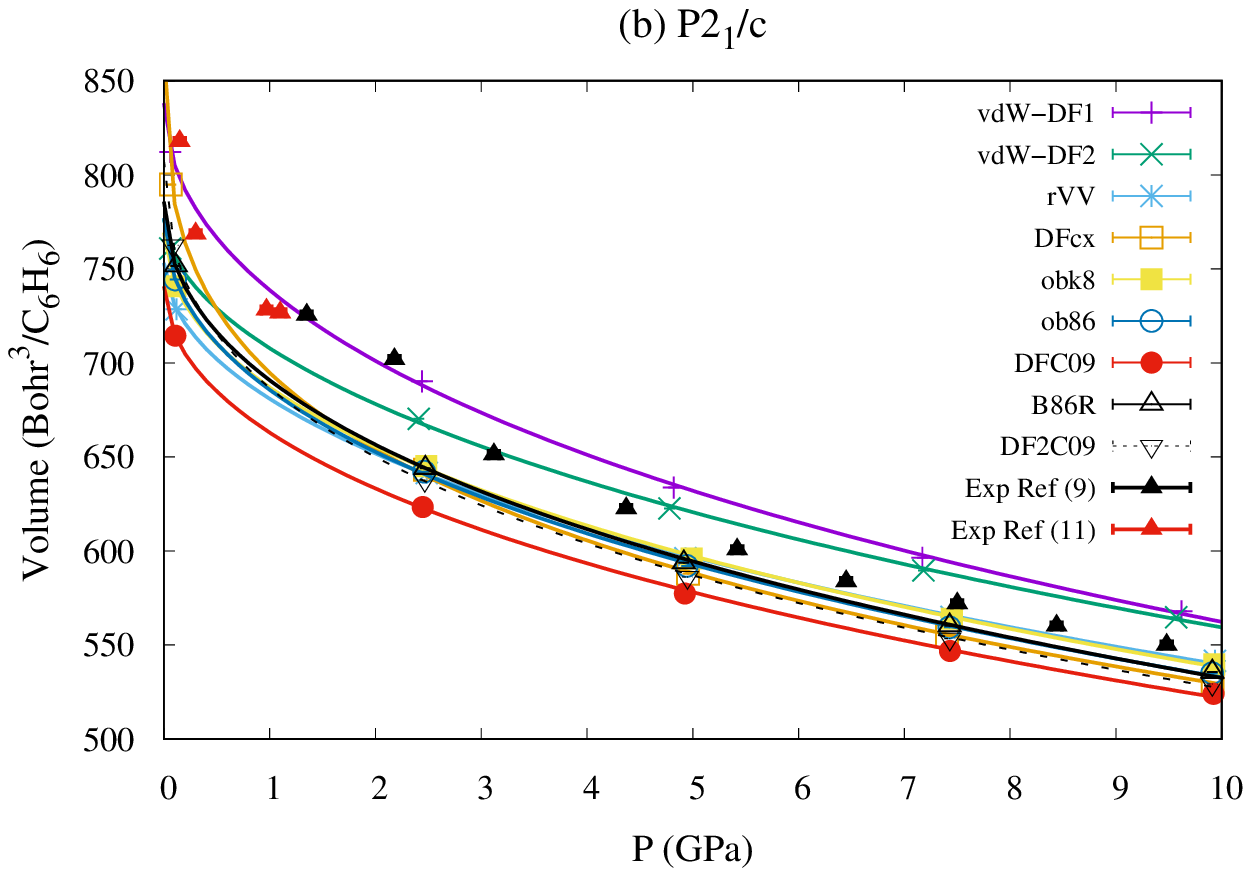} \\
\includegraphics[width=0.5\textwidth]{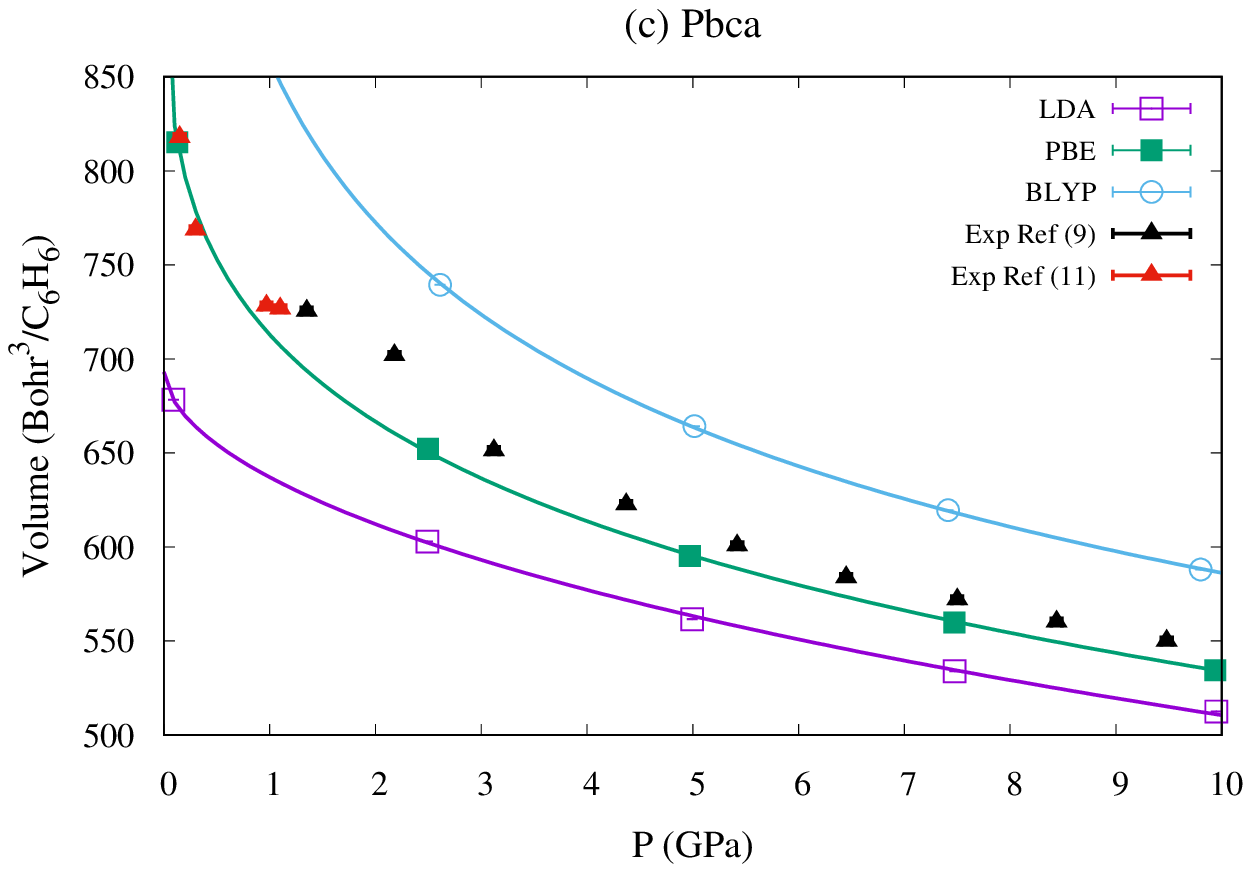}&
\includegraphics[width=0.5\textwidth]{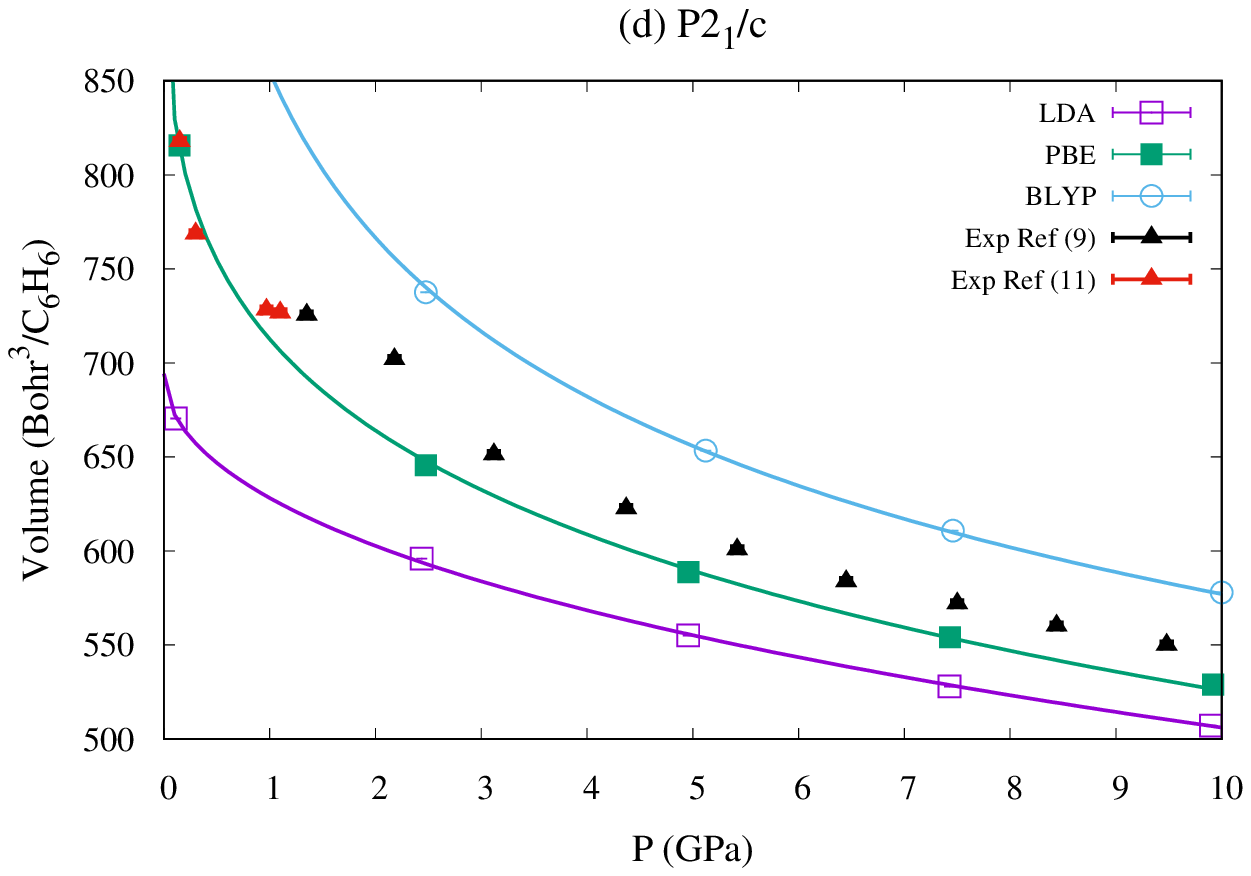} \\
\end{tabular}
\caption{\label{PV}(colour online) EOS of $Pbca$ and $P2_1/c$ structures 
obtained by vdW ((a) and (b)) and conventional ((c) and (d)) DFT functionals. 
The results are compared with experimental data which are reported in 
References \onlinecite{Ciabini1,Katrusiak2}.} 
\end{figure}

The vdW functionals yield different I$-$II phase transition pressure.
Figure~\ref{Ptran} illustrates $Pbca$ to $P2_1/c$ phase transition 
pressures which are obtained by different vdW functionals. The Cx\cite{cx}
and DF2C09\cite{C09} functionals show the lowest and highest phase 
transition pressures, respectively. The difference between largest and smallest 
phase transition pressure is about 1.1 GPa. This value corresponds to inaccuracy 
in prediction of phase transition pressure by vdW functionals. It should be noted that the 
experimental $Pbca$ to $P2_1/c$ phase transition occurs within 1.4 GPa pressure window.   
The results of the PBE, LDA, and BLYP functionals 
predict that the phase I$-$II transition occurs at 5.2, 5.6, and 3.5 GPa, respectively. 

Our phase diagram calculations indicate that vdW results are in better
agreement with experiments than the conventional functionals. 
Between the PBE and BLYP functionals,
the PBE $\delta_{C-C}$ results are closer to vdW-DF1 and vdW-DF2 $\delta_{C-C}$ at low pressures. 
The difference between PBE phase transition pressure 
and vdW-DF1 and vdW-DF2 phase transition pressures are 4.55 and 3.9 GPa,
respectively. However the difference between BLYP phase transition pressure 
and vdW-DF1 and vdW-DF2 phase transition pressures  are 2.85 and 2.2 GPa, respectively.    
As we discussed in the previous section, vdW-DF1 and vdW-DF2 $\delta_{C-C}$ are positive 
for both $Pbca$ and $P2_1/C$ structures below 2 GPa where the phase transition 
between them happens. Therefore the phase I$-$II transition in crystalline benzene 
occurs without any intermolecular contacts. This transition occurs only due to 
dispersion effects. 

Using our DFT results we compute the EOS of $Pbca$ and $P2_1/c$
structures. Figure~\ref{PV} represents the results which are obtained 
by vdW and conventional functionals. We compare our DFT results with experiments
which are reported in References \onlinecite{Ciabini1,Katrusiak2}.
The experimental results in Ref.~\onlinecite{Ciabini1}
are $V(P)$ data for crystalline benzene at 540 K that have been fitted 
by the Vinet EOS. The second experimental results~\cite{Katrusiak2}
belong to crystalline benzene at lower pressures and 295 K.
Among DFT conventional functionals used in this study only the PBE 
$V(P)$ results are close to experiments. The BLYP and LDA curves 
lie far above and below experimental curves, respectively. 
In general, the vdW results are in good agreement with experiments. 
At lower pressures vdW-DF1\cite{vdW1}, vdW-DF2\cite{vdW2}, and DFcx\cite{cx}
 $V(P)$ points for $Pbca$ phase 
are close to experiments. With increasing the pressure, the $P2_1/c$ $V(P)$ 
curves computed with vdW-DF2\cite{vdW2}, obk8\cite{Klimes1,Klimes2}, rVV\cite{rVV1,rVV2}, 
and B86R\cite{Klimes1,Klimes2} are close to 
experimental points. The rVV functional has a different nonlocal 
correlation kernel, whereas other vdW functionals are the modified versions of vdW-DF1 or vdW-DF2. 
Our EOS calculations indicate that the modifications bring the vdW-DF1 and vdW-DF2 $V(P)$ curves 
below experimental ones. It is hard to conclude whether these modifications improve 
the accuracy of vdW-DF1 and vdW-DF2 fucntionals, especially in the case of vdW-DF2 fucntional,
which overall gives the most accurate results. Our ground state EOS calculations indicate that 
at fixed pressure the volume per benzene molecule for $Pbca$ phase is larger than 
$P2_1/c$. This is in agreement with finite temperature experimental measurements. 
This conclusion is also independent of used DFT functionals.          

\begin{figure}
\begin{tabular}{c c}
\includegraphics[width=0.5\textwidth]{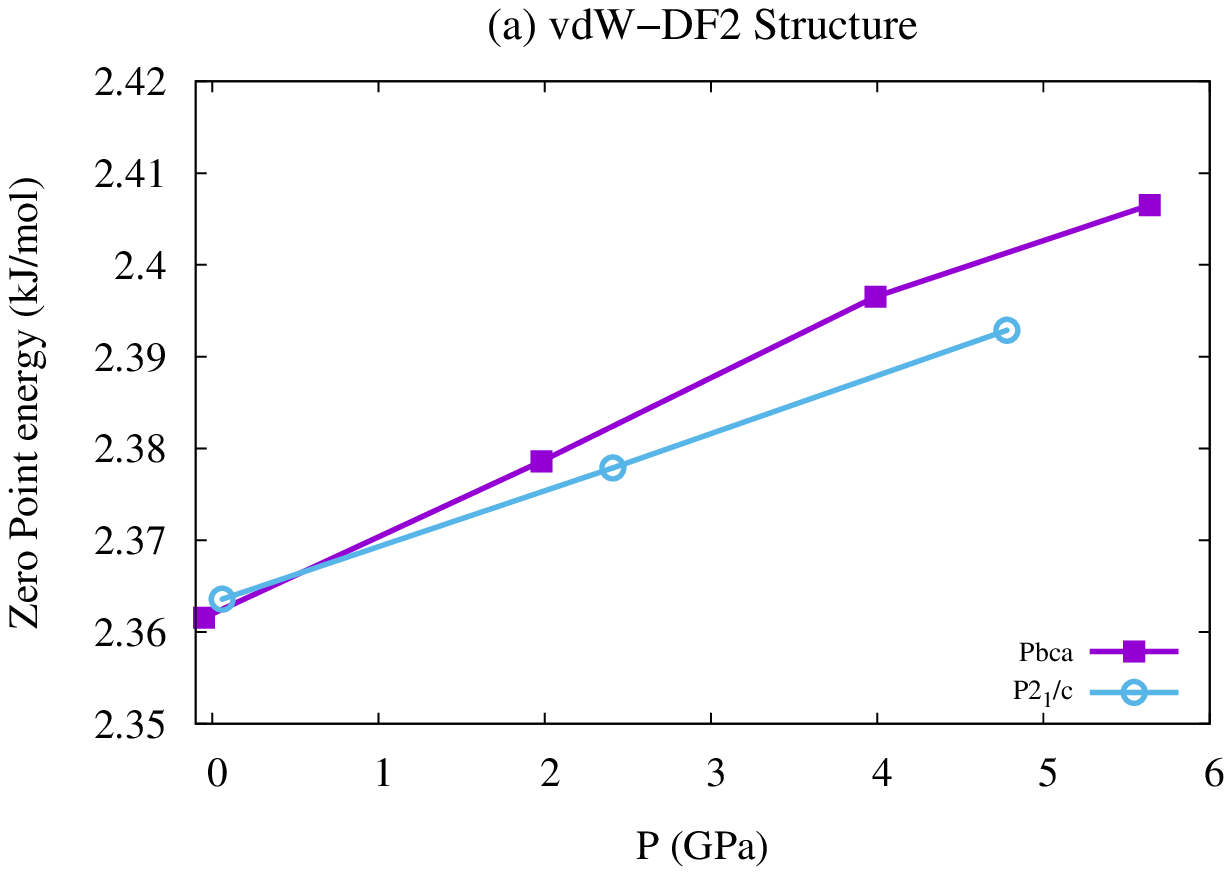}&
\includegraphics[width=0.5\textwidth]{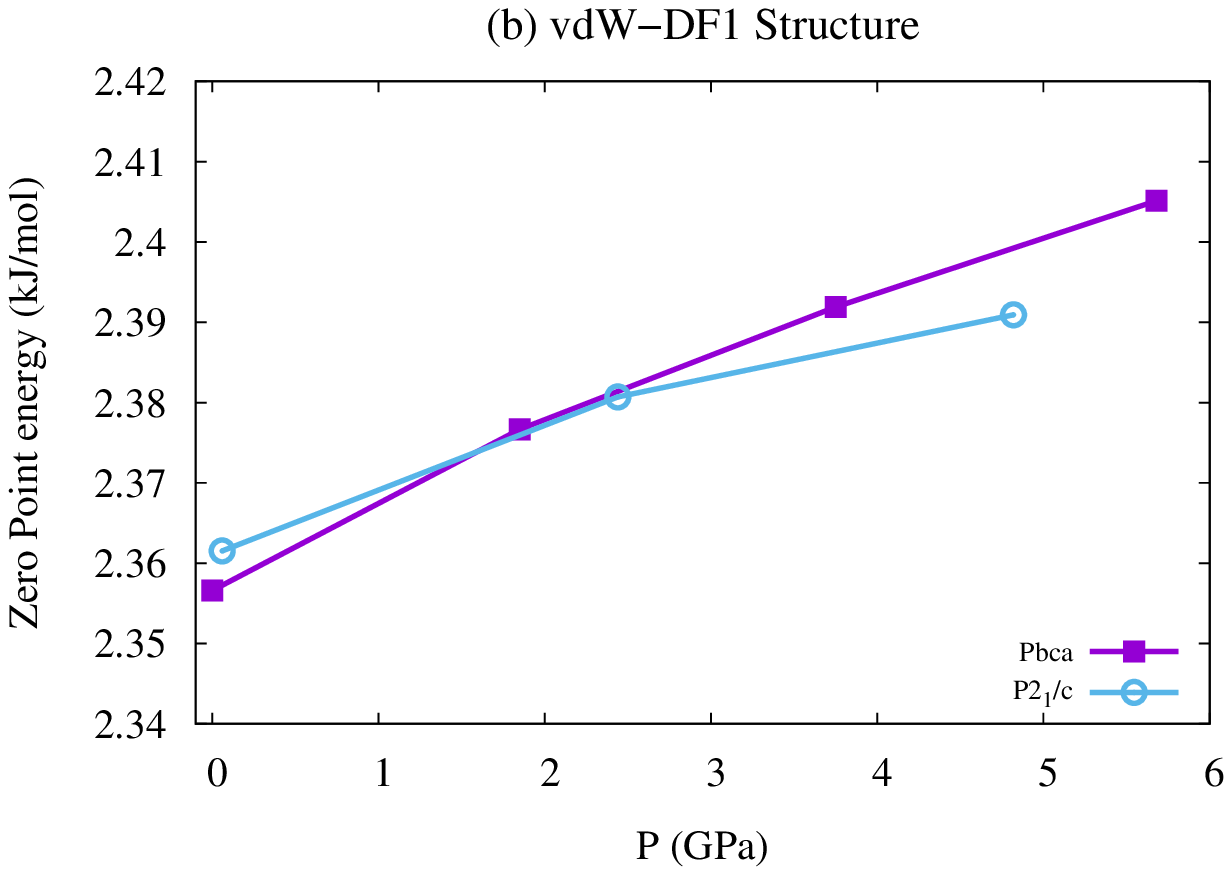} \\
\end{tabular}
\caption{\label{ZPE}(colour online) ZPE of $Pbca$ and $P2_1/c$ structures 
obtained by DFT. Geometries are accurately optimised by two functionals:
(a) vdW-DF2\cite{vdW2} and (b) vdW-DF1\cite{vdW1}.} 
\end{figure}

To investigate the ZPE contribution in phase diagram calculations, 
we simulated the difference between the gas and crystal ZPEs.  
The ZPE of the $Pbca$ and $P2_1/c$ structures with respect to gas phase 
is shown as function of pressure (Figure~\ref{ZPE}).  
We used the vdW-DF2 and vdW-DF1 functionals to optimise the structures for 
phonon calculations. ZPE is obtained using quasi-harmonic approximation, as 
explained in the previous section. Within the studied pressure range, 
the difference between the ZPE of 
phases I and II is less than 2 meV/atom.
The vdW-DF2 results indicate that the phase I$-$II ZPE transition
happens at 0.6 GPa, whereas the vdW-DF1 results predict that the 
phase I$-$II ZPE transition occurs at 1.65 GPa (Figure ~\ref{ZPE}).
The difference between the ZPE of phases I and II increases with pressure.
\begin{figure}
\begin{tabular}{c c }
 \includegraphics[width=0.5\textwidth]{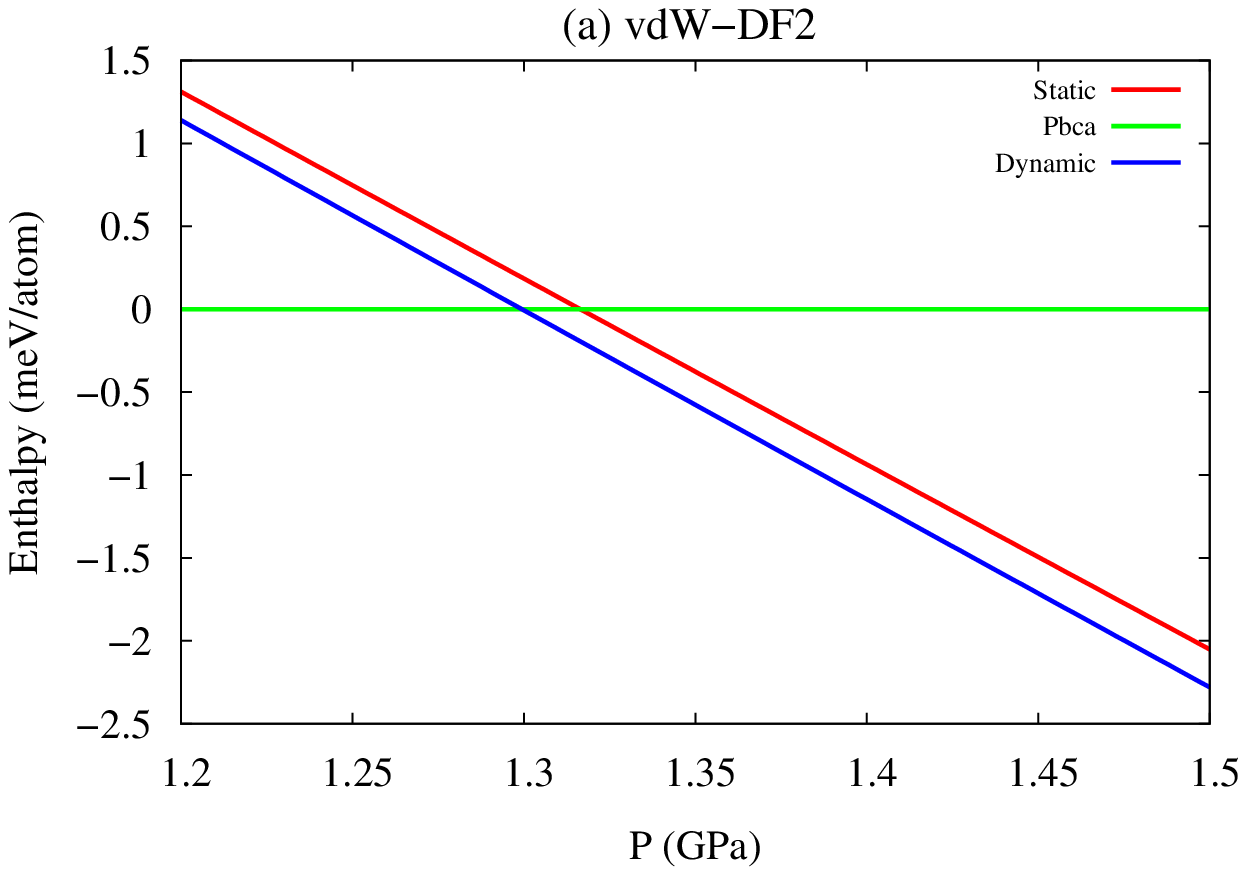}&
 \includegraphics[width=0.5\textwidth]{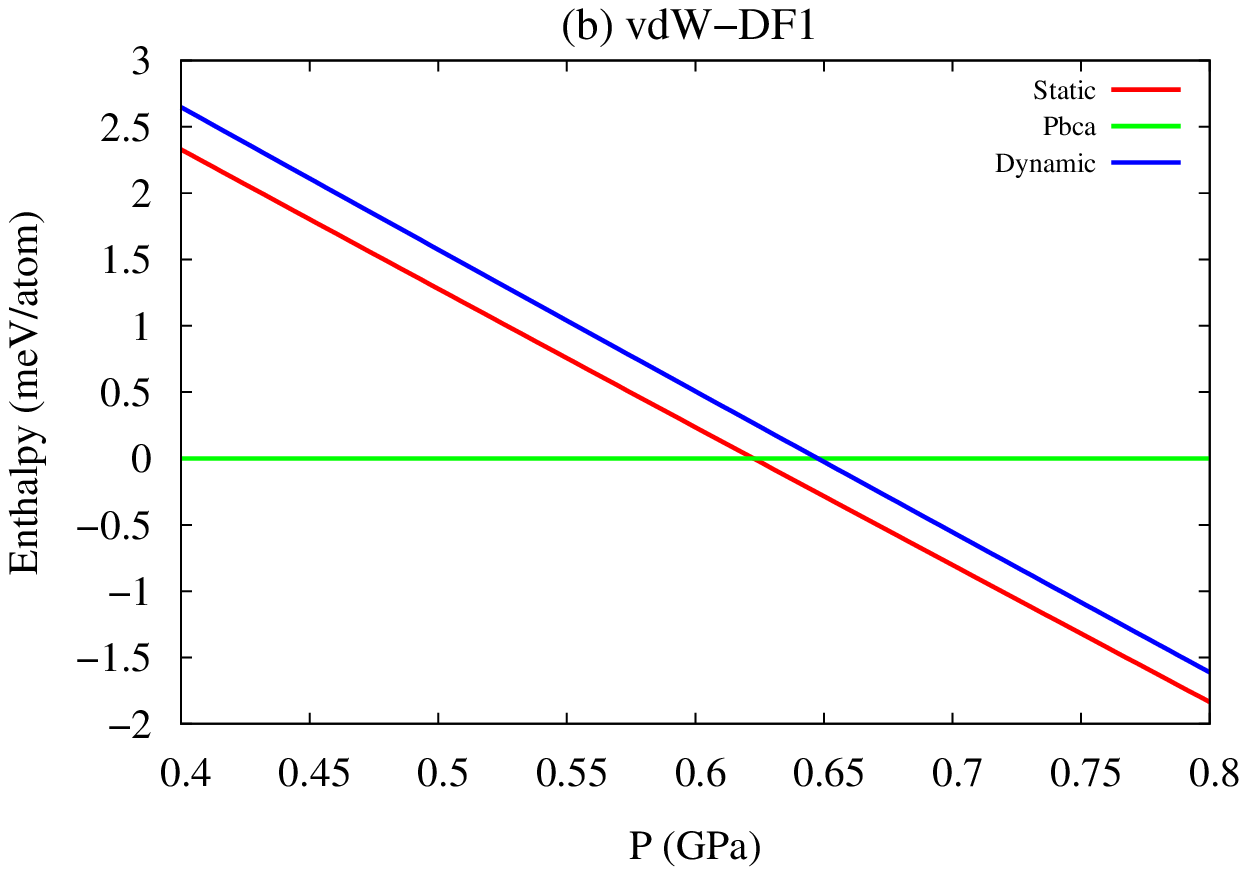}\\
\end{tabular}
\caption{\label{H_dyn}(colour online) Static and 
dynamic phase transition of $Pbca$ to $P2_1/c$ obtained by
(a) vdW-DF2\cite{vdW2} and (b) vdW-DF1\cite{vdW1} vdW functionals.} 
\end{figure}
The ZPE correction to the cohesive energy of crystalline benzene 
was previously calculated\cite{Bludsky}.
They evaluated the ZPE using $\Gamma$-point harmonic frequencies 
at the PBE level. They found that the ZPE of the $Pbca$ 
structure is 44 meV/molecule. In their calculations, they employed
experimentally reported\cite{David} orthorhombic cell without full 
three-dimensional optimisation.
Finite-temperature experiments\cite{Nakamura} show that 
the ZPE of crystalline benzene is 2.8 kJ/mol (29.02 meV/molecule).
The ZPE experimental result is also employed
to investigate the binding energy of benzene crystal\cite{Lu}. 
An estimate of 4.8 kJ/mol was obtained using DFT many-body dispersion
 method\cite{Reilly}. This ZPE is significantly larger than an estimate of 
2.8 kJ/mol which is obtained by finite molecular cluster 
calculations\cite{Podeszwa,Ringer}. Our ZPE results are close to 
PBC-DFT calculations\cite{Otero}, where an estimate of 2.6 kJ/mol is obtained 
using the PBE functional.    

The static phase diagrams in Figure \ref{DFT_enthalpy} 
assume that the atoms are infinitely massive. 
We calculate the dynamic phase diagram by adding the 
ZPE to the static results. Figure \ref{H_dyn} illustrates the dynamic 
phase diagrams of crystalline benzene at the DFT level. 
The vdW-DF2 results indicate that adding ZPE lowers the phase transition 
by 0.02 GPa, and the $Pbca$ to $P2_1/c$ phase transition pressure is 1.42 GPa.
The vdW-DF1 results predict that the phase transition 
occurs at 0.66 GPa, which is 0.03 GPa higher than the static phase 
transition pressure. 
The results of comparing the static and dynamic phase diagrams indicates that 
the ZPE contribution is negligible.

\subsection{Finite Temperature DMC Phase Diagram}
In this section we present our finite temperature phase 
diagram calculations. We use QMC based methods to calculate 
the electronic structure ground state energy.
The inadequacy of mean-field-like DFT calculations of hydrogen-rich 
systems was demonstrated before\cite{AzadiPRB}. To obtain reliable 
results, going beyond DFT-based methods and properly considering
many-body effects are necessary. The DMC is generally considered 
as the most accurate first-principle method available in studying the 
phase diagram of hydrogen-dominant materials\cite{Neil,Azadi3}.   
In addition DMC is an effective method to study non-covalent 
systems. It can reach and go beyond the chemical accuracy 
which is desired for non-covalent systems\cite{JCP15}. 

We perform DMC calculations to obtain  the wave-function-based
phase diagram for crystalline benzene at low-pressures.
We use the vdW-DF2 optimized structure for our DMC calculations. 
As we demonstrated in our DFT calculations, vdW-DF2 functional gives 
the closest results to experiment. 
The DMC results for energies in the limit of infinite system size
are obtained by extrapolation using DMC energy data at $1\times1\times1$
and $2\times2\times2$ simulation cells. Extrapolation is advantageous
because it can approximately account for finite-size effects that are not
considered in the other correction schemes, such as finite-size effects 
in the fixed-node error. In addition, it does not suffer from the 
reliance on stochastically optimised trial wave functions that affects  
the kinetic-energy correction, because it is purely based on SJ
DMC energies\cite{Azadi1,NeilFS}.

\begin{table}[h]
\caption{\label{Pbca_dmc} DMC energies of the $Pbca$ (phase I) structure.
Energies are obtained  at two simulation cells containing
$N_1$ = 48 and $N_2$ = 384 atoms. 
Linear extrapolated energies are shown as E($\infty$). Energy (E) and volume 
(Vol) are in eV and  Bohr$^3$ per benzene molecule, respectively.}
\begin{tabular}{ c c c c }
\hline\hline
    Vol  &  E($N_1$)    &  E($N_2$)     & E($\infty$) \\ \hline  
 781.086& -1024.7464(4)& -1022.7063(8) & -1022.4140(8)  \\ 
 693.335& -1024.6376(5)& -1022.5976(7) & -1022.3052(7)  \\ 
 646.304& -1024.4880(5)& -1022.4478(5) & -1022.1556(5)   \\
 619.413& -1024.3588(4)& -1022.3185(5) & -1022.0264(5)   \\
\hline\hline
\end{tabular}
\end{table}

Table ~\ref{Pbca_dmc} lists the DMC energies of the $Pbca$ structure at 
four primitive unit-cell volumes.
We consider two simulation cells for each density 
containing 48 and 384 atoms. DMC energy at thermodynamic  
limit is obtained by linear extrapolation in $1/N$. 

\begin{table}[h]
\caption{\label{P21c_dmc} DMC energies of the $P2_1/c$ (phase II) structure.
Energies are obtained  at two  simulation cells containing
 $N_1$ = 24 and $N_2$ = 192  atoms. 
Linear extrapolated energies are shown as E($\infty$). Energy (E) and volume 
(Vol) are in eV and  Bohr$^3$ per benzene molecule, respectively.}
\begin{tabular}{ c c c c }
\hline\hline
  Vol    &   E($N_1$)    &  E($N_2$)    & E($\infty$) \\ \hline 
 760.8398& -1024.8824(5) & -1022.6683(8)& -1022.3514(8)   \\ 
 670.2382& -1024.7736(6) & -1022.5582(8)& -1022.2426(8)   \\ 
 622.5712& -1024.6240(5) & -1022.4099(7)& -1022.0929(8)    \\
 589.6330& -1024.4948(5) & -1022.2793(8)& -1021.9637(8)    \\
\hline\hline
\end{tabular}
\end{table}

Table ~\ref{P21c_dmc} shows the DMC energies of the $P2_1/c$
 structure at different primitive unit-cell volumes.
We consider two simulation cells for each density 
containing 24 and 192 atoms. DMC energy at infinite system 
size limit is calculated by linear extrapolation in $1/N$. 

To identify enthalpy-pressure curves for the $Pbca$ and $P2_1/c$
structures, we fitted model equations of state $E(V)$ to our 
finite-size-corrected DMC energy against volume $V$.
We used the Vinet EOS\cite{Vinet} to fit our 
total energies and propagate errors using classical statistics. 
The pressure $P = -(\partial E/\partial V)$
and the enthalpy is $H=E+PV$, where 
$E$ is DMC electronic structure energy of system.

\begin{figure}
\begin{tabular}{c c}
\includegraphics[width=0.5\textwidth]{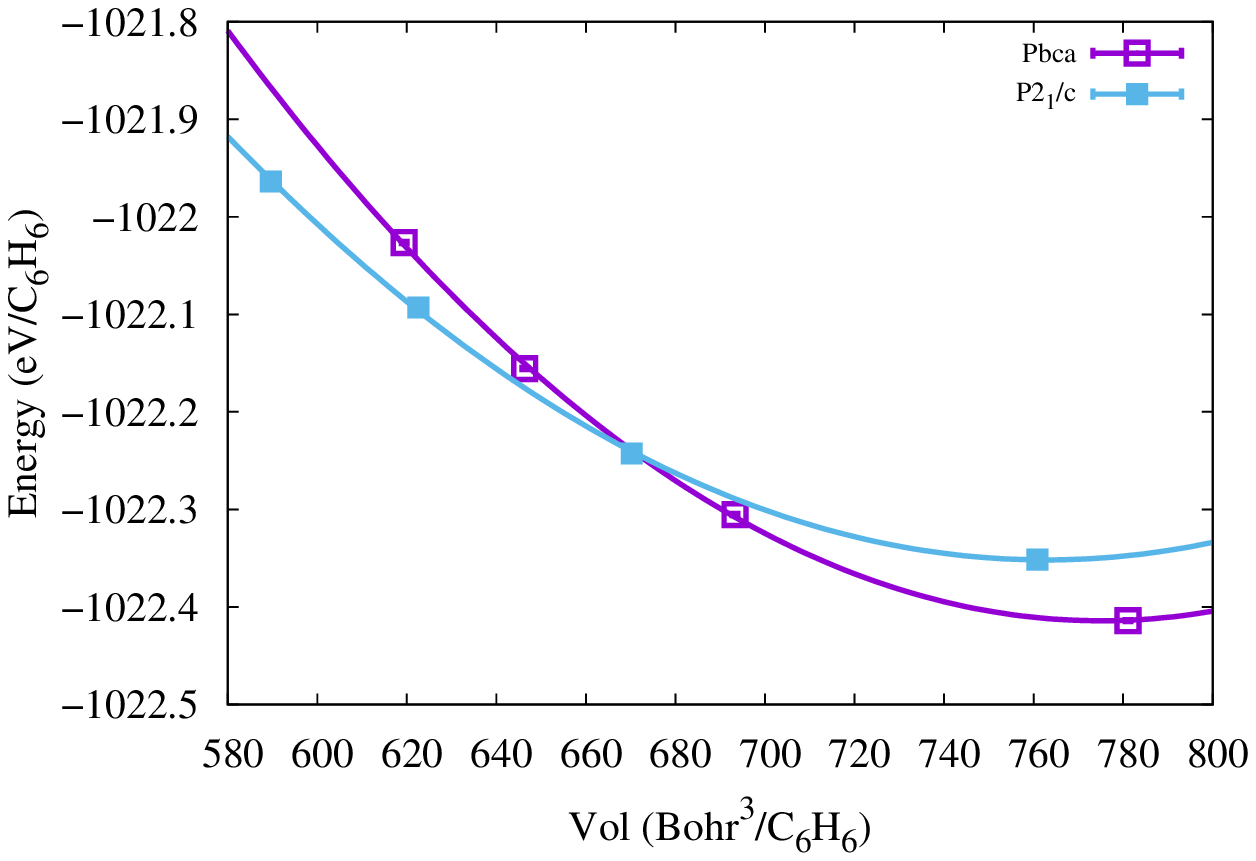}&
\includegraphics[width=0.5\textwidth]{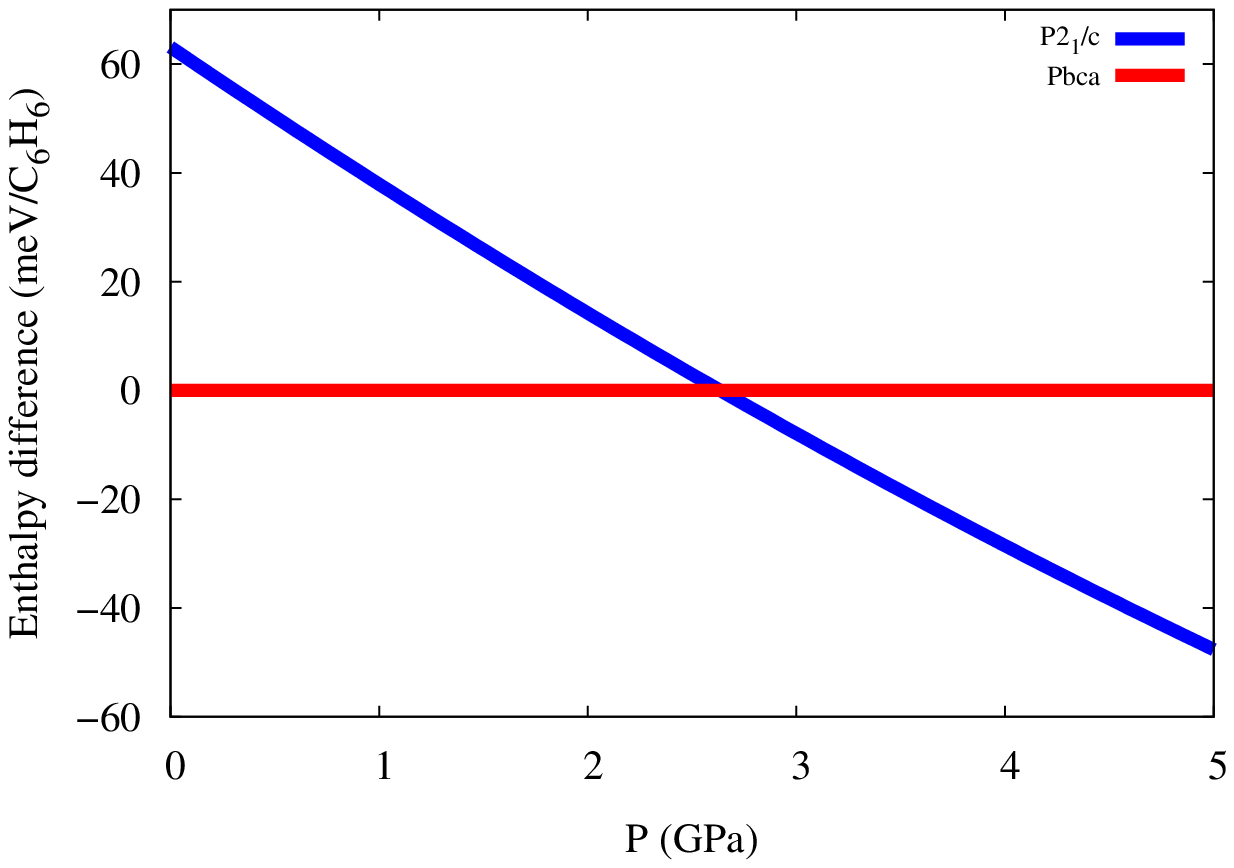}
\end{tabular}
\caption{\label{E_qmc}(colour online) (left) DMC energy of the $Pbca$ and 
$P2_1/c$ structures as function of volume per benzene molecule. 
Energy error bars are included in point sizes and are of the 
order of meV. (right) Relative enthalpies of the $Pbca$ and 
$P2_1/c$ structures as function of pressure. The widths of 
the DMC lines indicate the estimated uncertainties
in the enthalpies because of statistical and systematic errors.}  
\end{figure}

Figure~\ref{E_qmc} (left) illustrates the DMC energy of phases I and II 
of crystalline benzene as function of volume per benzene molecule. 
With increasing density, phase II becomes favourable
over phase I in the $Pbca$ structure. Figure~\ref{E_qmc} (right) shows 
the relative enthalpies of the $Pbca$ and $P2_1/c$ structures.
Based on our static enthalpy-pressure phase diagram, the $Pbca$ to 
$P2_1/c$ phase transition occurs at pressure 2.6$\pm$0.1 GPa. 
The use of the DMC method has significant consequences for the 
static-lattice relative enthalpies of the studied structures.
Compared with vdW-DF2, the DMC enthalpy-pressure results predict that 
the phase I$-$II transition occurs at 1.2 GPa higher pressure.  
Among conventional DFT functionals, the BLYP results are closest to DMC. 
The difference between DMC and BLYP phase transition pressure is 
0.9 GPa.  

\begin{figure}
 \begin{tabular}{c c}
\includegraphics[width=0.5\textwidth]{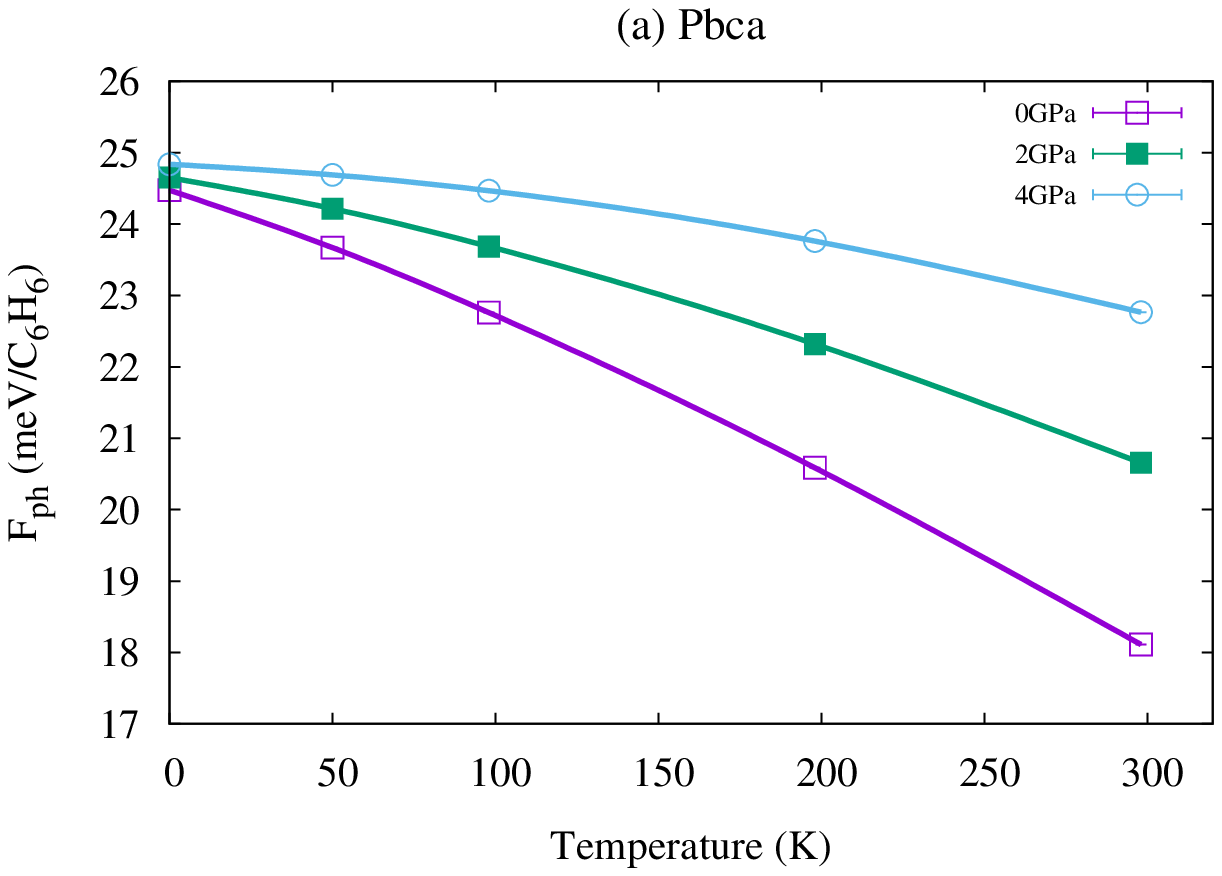}&
\includegraphics[width=0.5\textwidth]{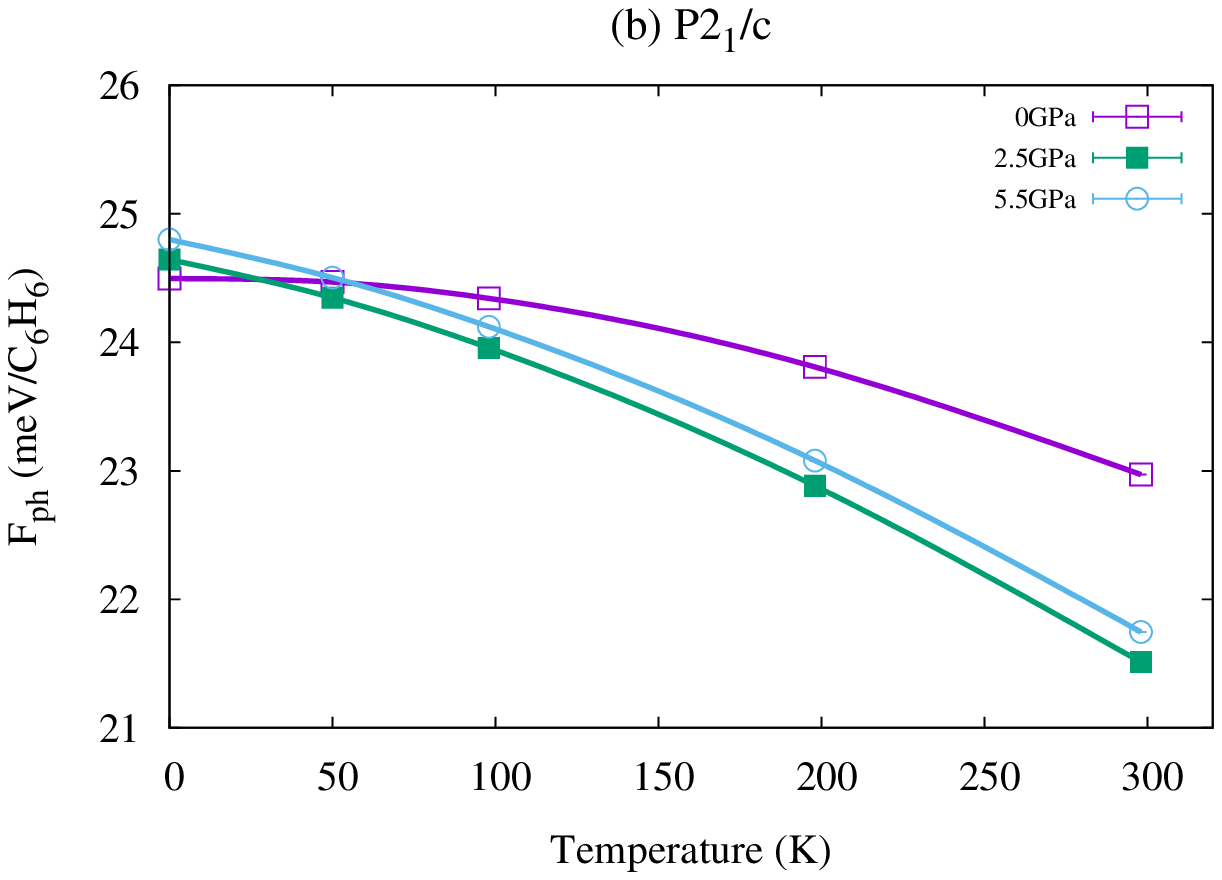}\\
 \end{tabular}
\caption{\label{Fvib} Phonon contribution to the Helmholtz free energies $F_{ph}$ of 
the $Pbca$ and $P2_1/c$ structures of crystalline benzene. The geometries are optimized  
using vdW-DF2\cite{vdW2} functionals.}  
\end{figure}

To obtain the phase diagram at finite temperature, 
we used quasi-harmonic approximation to obtain lattice 
dynamic contribution to the free energies.
Phonons have contributed to the Helmholtz free energies
$F_{ph}$ of crystalline benzene (Figure~\ref{Fvib}).
We used vdW-DF2
functionals to optimise the $Pbca$ and $P2_1/c$ structures at different 
pressures. Vibrational free energies are calculated at different temperatures 
of 50, 100, 200, and 300 K. At room temperature and 0 GPa vibrational free
energy of $P2_1/c$ is higher than $Pbca$. 
Meanwhile, the vibrational free energies of $Pbca$ become
higher than $P2_1/c$ by increasing the pressure.
This results indicates the stability of the $Pbca$ phase at ambient conditions, 
which is also observed experimentally\cite{Ciabini1,Ciabini2}.   

\begin{figure}
 \begin{tabular}{c c}
      \includegraphics[width=0.5\textwidth]{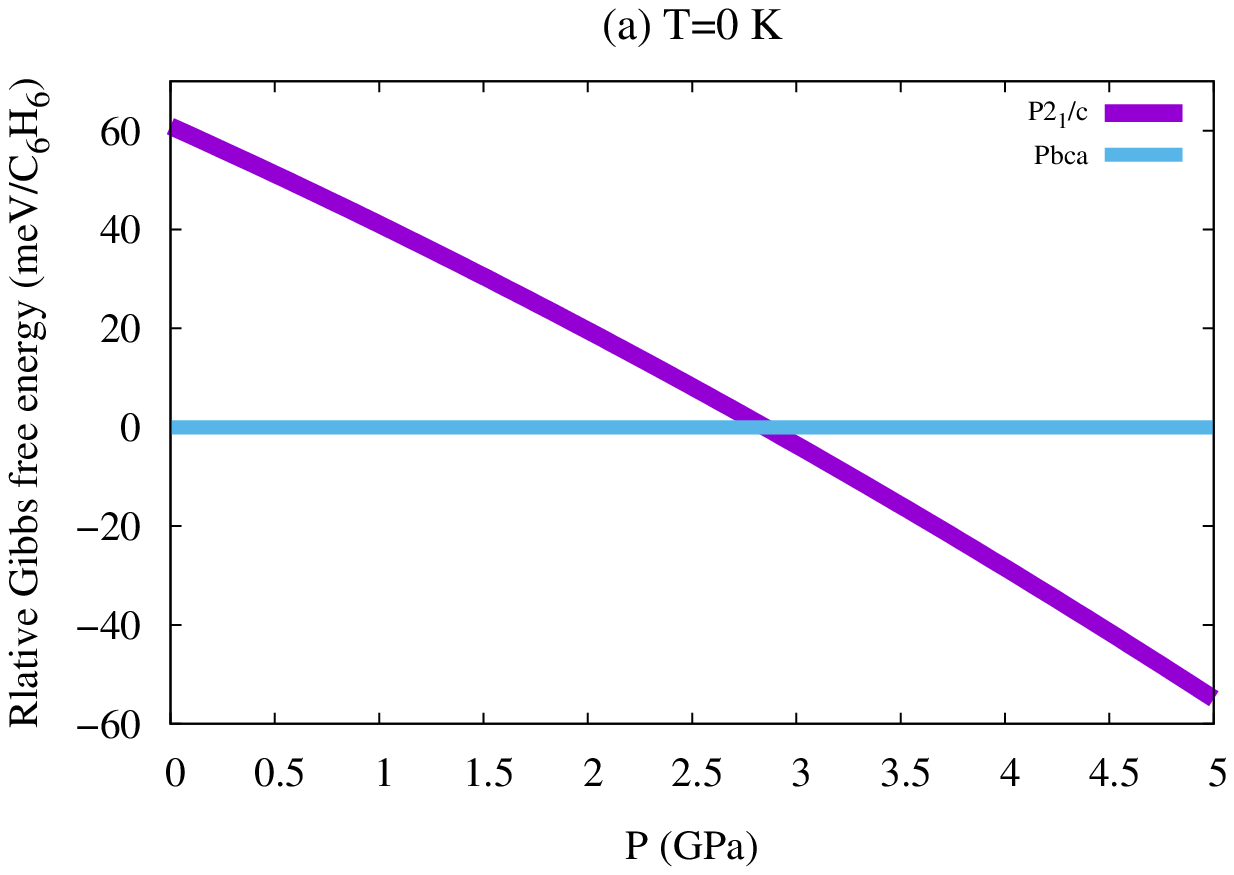}&
      \includegraphics[width=0.5\textwidth]{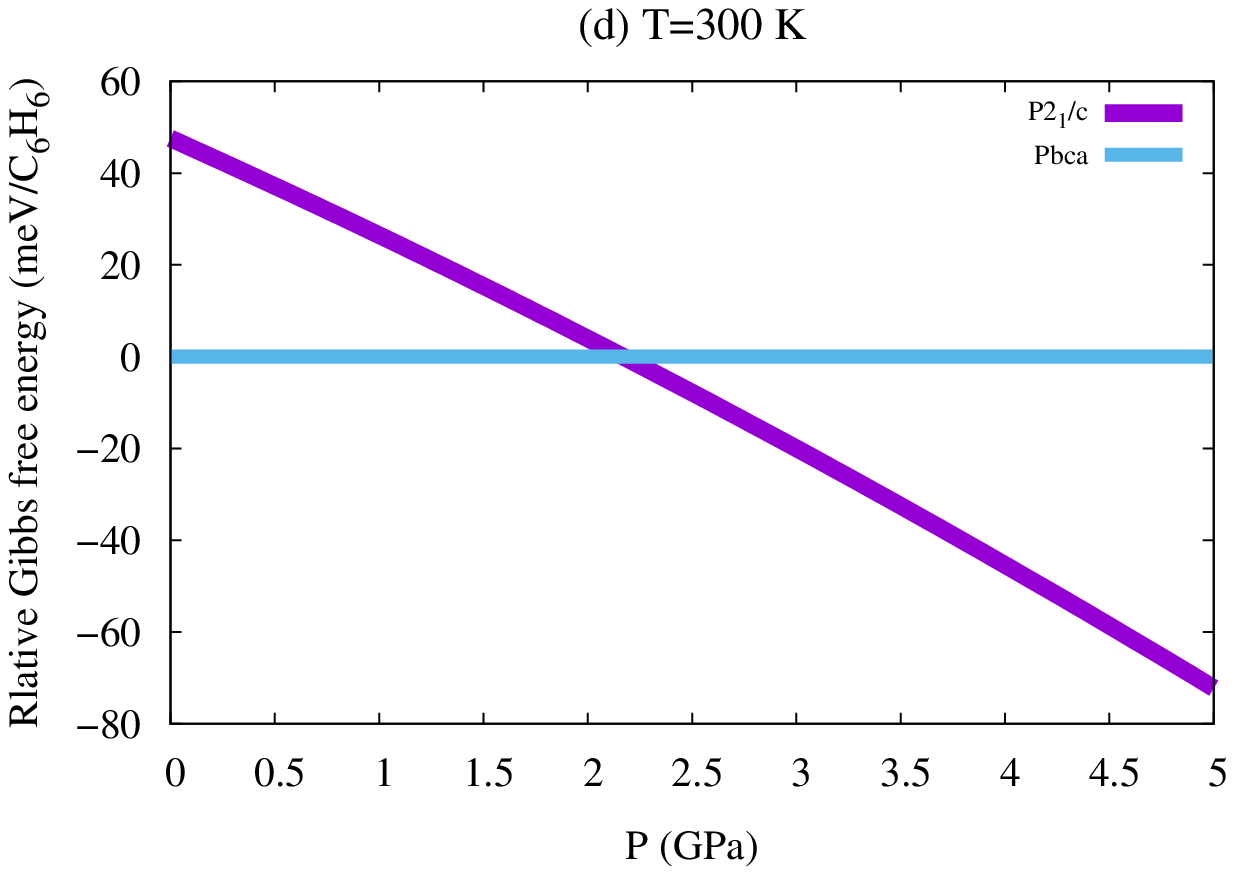}\\
 \end{tabular}
\caption{\label{Gibbs} Relative Gibbs free energies of the $Pbca$ and $P2_1/ca$ 
structures. (a) 0 K, and (d) 300 K. The Gibbs free energies are
calculated using static-lattice DMC calculations together with DFT quasi-harmonic 
vibrational calculations.}  
\end{figure}

We calculated relative Gibbs free energies 
of the $Pbca$ and $P2_1/c$ structures at different temperatures (Figure~\ref{Gibbs}).
The static electronic structure results are obtained by DMC calculations.
Our results predict that the room temperature $Pbca$ to $P2_1/c$ 
structure transformation happens at 2.1(1) GPa.
Experiments indicate the transition to phase II occurs
at room temperature and around 1.4 GPa\cite{Piermarini}. 
The I to II phase transition was found 
to be extremely sluggish, and it can be speeded up by heating 
the sample\cite{Thiery}. 
Keeping the low-pressure phase I, $Pbca$, in a metastable state
at least up to 3 GPa is possible without heating\cite{Bridgman}.
 Experimentally achieving low-enough temperature results is extremely difficult.
Our DMC phase diagram at low temperature predicts that the $Pbca$ phase 
could be stable up to 2.6(1) GPa.   
The phase diagram that we obtained by combining DMC static-lattice
energies and quasi-harmonic vibrational energies can be extended 
to higher pressures. 

As the final step of our study,
we calculated the lattice energy of crystalline benzene at ambient conditions.
The cohesive energy yields the strength of the vdW forces holding the crystalline 
benzene together.
We used our DMC and ZP energies for $Pbca$ structure. 
The cohesive energy is calculated using the difference between total energies of 
$Pbca$ structure and its fragments. Cohesive energy calculation is a precise test 
of DMC method, since it has to accurately describe two different systems of benzene 
molecule and crystalline benzene. The electronic structure of these two systems are 
not similar. In our DMC lattice energy calculation, we used 
same time step of 0.01 a.u for both crystal and molecule.  
We found an estimate of 
50.6$\pm$0.5 kJ/mol for lattice energy. Ab initio many-electron wave functions 
methods provide an estimate of 55.90$\pm$0.76 kJ/mol for benzene crystal lattice
energy at zero temperature\cite{Yang}. The experimental lattice energy at same 
condition is 55.3$\pm$2.2 kJ/mol\cite{Yang}. 
We used conventional Jastrow factor in our DMC 
calculations. In principle, the DMC lattice energy can be systematically improved
by accurately taking into account the correlation energy and also decreasing 
the fixed-node errors. These purposes can be fulfilled  
by adding additional terms in Jastrow factor and using backflow  
transformations\cite{JCP15}. However, improving the DMC lattice energy until 
it converges to exact results requires huge amount of computational time.   

\section{Conclusion}

We have comprehensively studied the crystalline benzene phase 
diagram at pressures below 10 GPa. We have used 
different vdW functionals and also three most used conventional 
functionals to obtain DFT energy of system. 
The vdW-DF2 results of our study indicated
that the  $Pbca$ and $P2_1/c$  structures are the 
best candidates for phases I and II, respectively. We have used 
the accurate DMC method to calculate the ground-state electronic structure 
energy of system. We have compared static enthalpy-pressure phase 
diagrams which are obtained by DFT and DMC methods. We used quasi-harmonic 
approximation and density functional perturbation theory
 to calculate the phonon contribution to the 
free energy of system. Our Gibbs free energy phase diagram predicts 
that at room temperature, the phase I$-$II transition occurs at 2.1(1) GPa,
which is in good agreement with experiments. We have found DMC lattice energy 
of 50.6$\pm$0.5 kJ/mol for crystalline benzene at ambient conditions. 
The results of our study indicate the importance of many-body electronic structure 
calculation to obtain a reliable phase diagram for molecular crystals.
 
\begin{acknowledgments}

This study utilised computing facilities provided by ARCHER,
the UK national super computing service, and by the University
College London high-performance computing centre.
S. Azadi acknowledges that the results of this research have been obtained using the
PRACE-3IP project (FP7 RI-312763) resource ARCHER based in the UK.
The authors acknowledge the financial support of the European Research Council 
under the Advanced Grant ToMCaT (Theory of Mantle, Core,
 and Technological Materials). R. E. Cohen 
acknowledges the support of the Carnegie Institution for Science 

\end{acknowledgments}

\end{document}